\documentclass[traditabstract]{aa}
\usepackage{natbib}
\usepackage{graphicx}
\usepackage{subfig}
\usepackage[cmex10]{amsmath}
\usepackage{amssymb}
\usepackage{alltt}
\usepackage{url}
\usepackage{txfonts}
\usepackage{color}
\renewcommand{\today}{January 11, 2012}

\title{A morphological algorithm for improving radio-frequency interference detection}
\author{A.R.~Offringa\inst{1} \and J.J.~van~de~Gronde\inst{2} \and J.B.T.M.~Roerdink\inst{2}}
\institute{University of Groningen, Kapteyn Astronomical Institute, PO Box 800, 9700 AV Groningen, The Netherlands.\\\email{offringa@astro.rug.nl} \and University of Groningen, Johann Bernoulli Institute for Mathematics and Computer Science, P.O. Box 407, 9700 AK Groningen, The Netherlands.\\\email{j.j.van.de.gronde@rug.nl} and \texttt{j.b.t.m.roerdink@rug.nl}}
\titlerunning{A morphological algorithm for improving RFI detection}
\date{Received 22 November 2011 / Accepted 11 January 2012}
\begin{document}%
\abstract{A technique is described that is used to improve the detection of radio-frequency interference in astronomical radio observatories. It is applied on a two-dimensional interference mask after regular detection in the time-frequency domain with existing techniques. The scale-invariant rank (SIR) operator is defined, which is a one-dimensional mathematical morphology technique that can be used to find adjacent intervals in the time or frequency domain that are likely to be affected by RFI. The technique might also be applicable in other areas in which morphological scale-invariant behaviour is desired, such as source detection. A new algorithm is described, that is shown to perform quite well, has linear time complexity and is fast enough to be applied in modern high resolution observatories. It is used in the default pipeline of the LOFAR observatory.}

\maketitle  

\section{Introduction}
Modern telescopes in radio astronomy, such as the Expanded Very Large Array (EVLA) and the Low-Frequency Array (LOFAR), are sensitive devices that observe the sky with enormous depth and detail. The observed bandwidth of telescopes has dramatically increased over the last decades, and often overlaps with parts of the radio spectrum that have not been reserved for radio astronomy. Simultaneously, the radio spectrum is becoming more crowded because of technological advancement. Therefore, radio observations are affected by man-made radio transmitters, which can be several orders of magnitude stronger than the weak celestial signals of interest. This kind of interference, which seriously disturbs radio observations, is called radio-frequency interference (RFI).

Numerous techniques have been suggested to perform the challenging task of RFI mitigation. They include using spatial information to null directions, provided in interferometers or multi-feed systems \citep{ellingson-spatial-nulling-2002, multichannel-rfi-mitigation, hampson-spatial-nulling-2002, spatial-filtering-parkes-multibeam}; removing the RFI by using reference antennas \citep{adaptive-cancellation}; and blanking out unlikely high values at high time resolutions \citep{wsrt-rfims, multichannel-rfi-mitigation, pulse-blanking, chi-square-time-blanking-weber}. Despite the numerous possible techniques, almost any observation needs to be post-processed due to RFI effects. The most used technique for such a final processing step consists of detecting the RFI in time, frequency and antenna space, and ignoring the contaminated data in further data processing. This step is often referred to as ``data flagging''. Historically, this step was performed by the astronomer. However, because of the major increase in resolution and bandwidth of observatories, leading to observations of tens of terabytes, this is no longer feasible. The tendency is therefore to implement automated RFI flagging pipelines in the observatory's pipeline. Examples of these are the RFI mitigation pipeline used for the Effelsberg Bonn HI Survey \citep{effelsberg-rfi-mitigation} and the AOFlagger pipeline \citep{LOFAR-RFI-pipeline}.

\subsection{RFI detection}
\begin{figure*}
 \begin{center}  
  \subfloat[]{ \label{fig:wsrt-example-unflagged}
   \includegraphics[width=70mm]{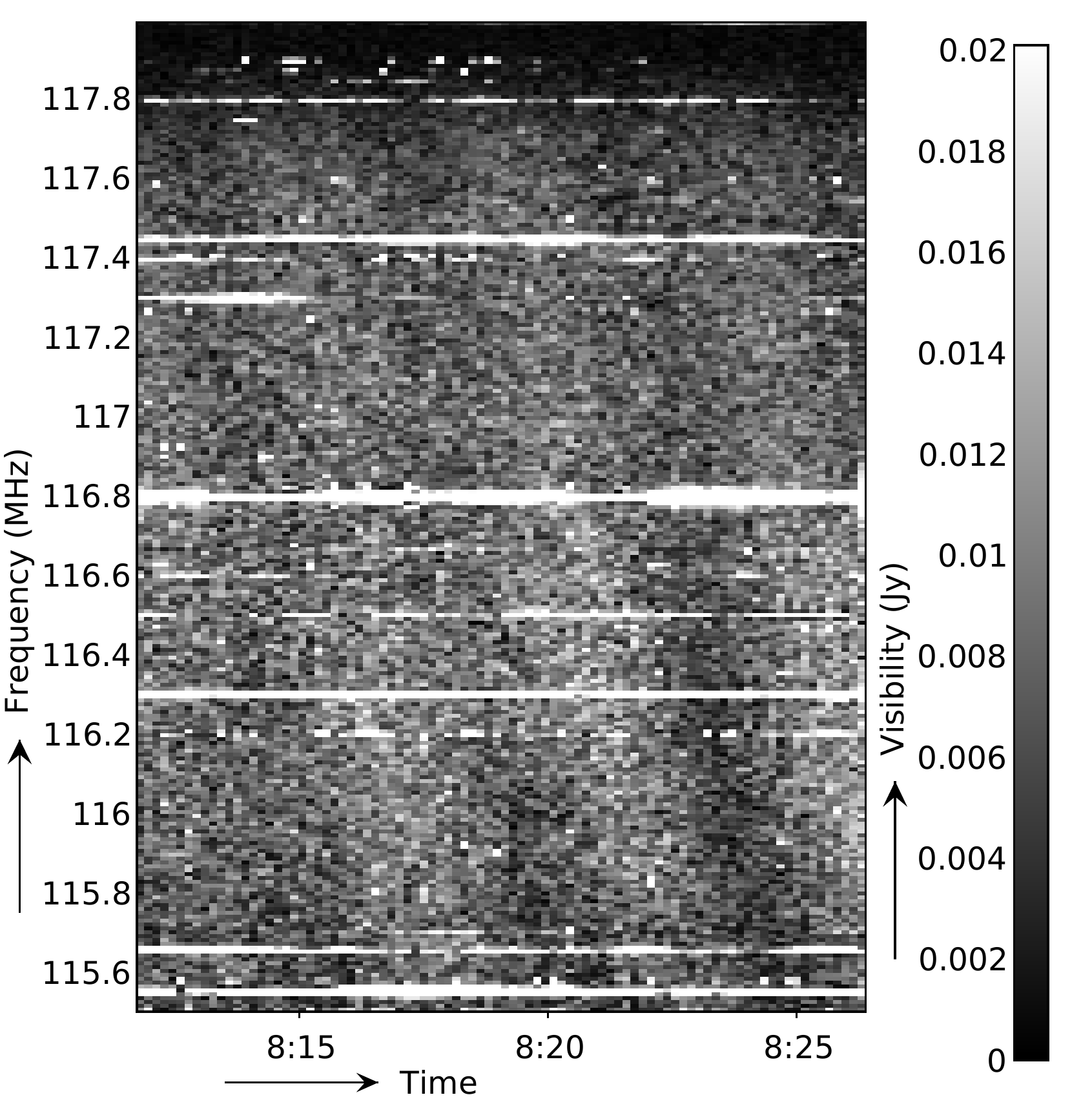}
  }\hspace{3mm}\nolinebreak%
  \subfloat[]{ \label{fig:wsrt-example-flagged}
   \includegraphics[width=70mm]{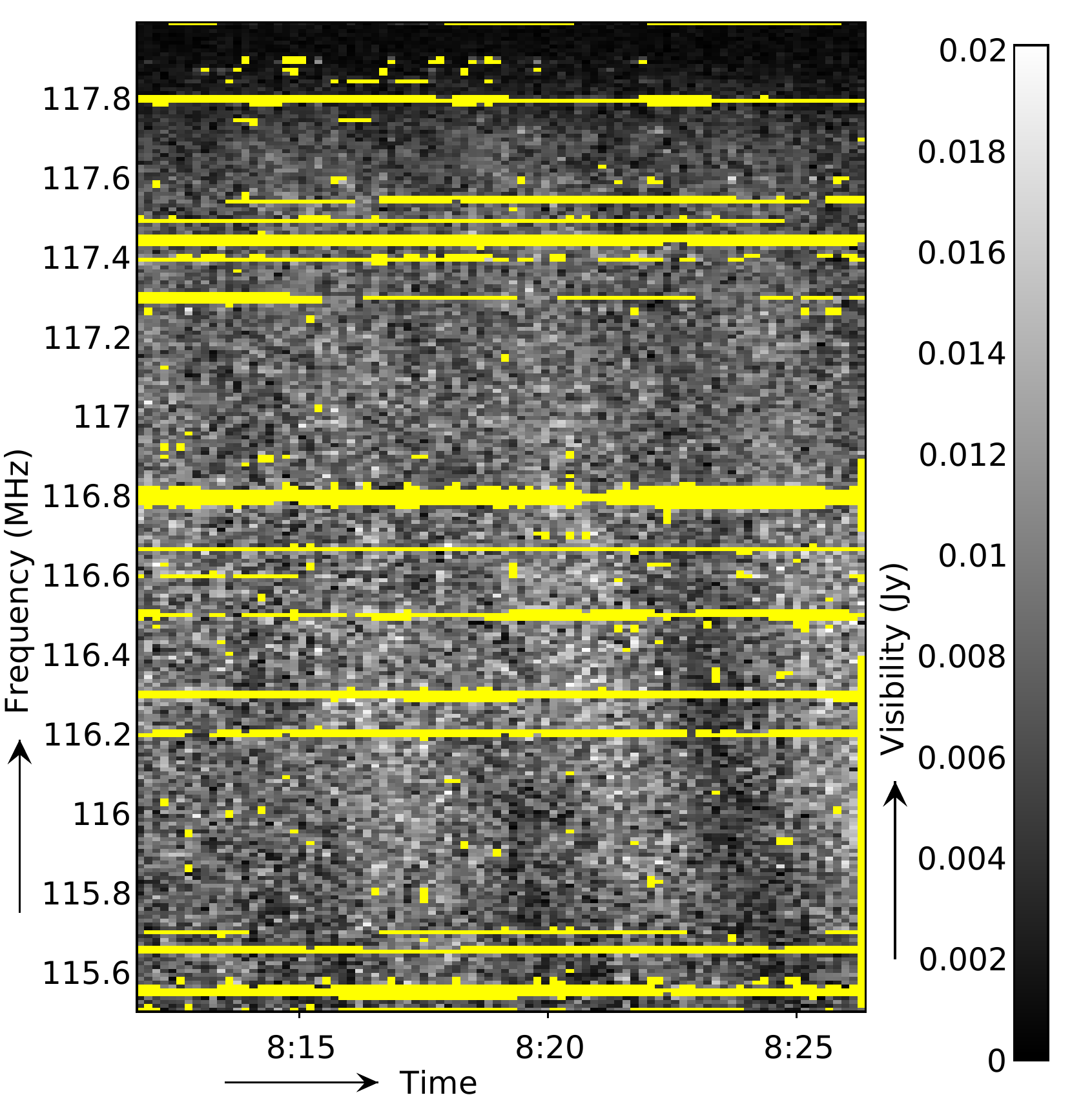}
   }%
 \end{center}
 \caption[]{Typical spectral line RFI received in a short period of WSRT data around 117 MHz. It is likely that such RFI sources transmit continuously within a small bandwidth. Panel~\subref{fig:wsrt-example-unflagged} shows the original observation, while panel~\subref{fig:wsrt-example-flagged} shows what the AOFlagger with default settings would flag without morphology-based flagging. Detection is quite accurate, but some of the detected lines in panel~\subref{fig:wsrt-example-flagged} are not continuous. It is likely that those RFI sources were active in the gaps as well. Morphology-based detection will help in such cases. The plot shows Stokes I amplitudes of the cross-correlation of antennas RT0 $\times$ RT1, which is a 144m East-West baseline. A single pixel is 10 seconds $\times$ 10 kHz of data.}
 \label{fig:wsrt-example}
\end{figure*}
RFI detection often involves thresholding based on amplitude \citep{statistical-rfi-removal, post-correlation-rfi-classification}, although also higher order statistics such as the kurtosis have been used \citep{hardware-implementation-spectral-kurtosis}. The latter requires storing both the mean powers and the squared powers, thereby doubling the data rate, and hence is not always usable. Most interfering sources radiate either in a constant small frequency range, or produce a broadband peak in a short time range. Examples of such interferences are respectively air traffic communication and lightning. Consequently, an interfering source tends to affect multiple neighbouring samples in the time-frequency domain. These samples form straight lines, parallel to the time and frequency axes. An example is given in Fig.~\ref{fig:wsrt-example}\subref{fig:wsrt-example-unflagged}, which shows data from the Westerbork Synthesis Radio Telescope (WSRT). This line-shaped behaviour of RFI can be used to improve the accuracy of detection algorithms. An algorithm that uses this information is the \texttt{SumThreshold} method, which shows a very high detection accuracy compared to other methods \citep{post-correlation-rfi-classification}. This method is used in one of the steps in the AOFlagger pipeline \citep{LOFAR-RFI-pipeline}. An important consideration for succesful application of automated feature detection algorithms such as these, is that the signal of interest should not contain significant line-shaped features, as is the case with spectral line observations. Also, methods that assume straight, one-dimensional features in the time-frequency domain, might not work well in situations where the features are curved. This can occur when both the frequency and the time resolutions are high enough to resolve frequency variation in sources, for example when sources are Doppler shifted or vary intrinsically, such as with certain radar signals. With LOFAR, we see very few such sources.

Typically, the received power of interfering sources varies over time and frequency. This happens because of several effects, such as intrinsic variation of the source; changing ionosphere; and because of instrumental effects. A typical example of the latter, which is present in almost every observation, is the change of the telescope's gain towards a terrestrial source as the telescope tracks a field in the sky. Like time variation, frequency variation can be caused intrinsically by the source. The instrument also adds frequency-dependent gain, for example due to imperfect band-pass filters. Even though a source might be continuously received by the telescope, thresholding detection methods might fail to detect the interferer over its full range due to the variation in received power. Figure~\ref{fig:wsrt-example}\subref{fig:wsrt-example-flagged} shows an example where this is likely the case. Increasing the sensitivity of the thresholding method might help somewhat, but will also cause an increase of false positives. While some falsely detected samples are tolerable, they should be kept minimal in order to avoid data bias and insufficient resolution.

Using mathematical morphology for RFI detection is not a new idea; a dilation is often used during RFI processing to flag areas near high values in the time-frequency domain. An example of this can be found in \citet{statistical-rfi-removal}, where windows of 5 time steps $\times$ 5 frequency channels around detected samples are flagged. However, standard morphological techniques are not scale invariant. An operator is called scale invariant if scaling its input results in the same scaling of its output. An ordinary dilation will cause sharp RFI features to create a high amount of false positives, while flagging smooth RFI features requires a very large dilation kernel. Another scale-dependent technique used for RFI detection is to consider the statistics of time steps and frequency channels. In this paper, we will show that scale invariance is a desirable property of RFI detection algorithms. This paper provides:
\begin{itemize}
 \item A detailed description of a recently-introduced morphological technique for RFI detection;
 \item Analysis of the technique and a comparison with an ordinary dilation, using simulations and real data from two different radio-observatories;
 \item A novel fast algorithm with linear time complexity to implement the technique.
\end{itemize}

\subsection{Outline}
We discuss a morphological technique that can be used to improve RFI detection. The method flags additional samples that are likely to be contaminated with RFI, based on the morphology of the flag mask output of a thresholding stage in the pipeline. In sect.~\ref{sec:technique-description} we describe the technique and show a fast algorithm to implement it. We present some results of the method on simulated data and real data in Sect.~\ref{sec:results}. Finally, we summarize and discuss the results in Sect.~\ref{sec:conclusions}.

\section{The scale-invariant rank operator} \label{sec:technique-description}
RFI features such as in Fig.~\ref{fig:gaussian-broadband}\subref{fig:gaussian-broadband-no-noise} are common in radio observations, and can occur at different scales. However, a morphological dilation is not \emph{scale invariant}, and will thus necessarily work better for some RFI features than others. To overcome this problem, we will describe and analyse a morphological rank operator that is scale invariant\footnote{The mathematical properties of this technique will be analysed in more detail in van de Gronde et al., 2012, in preparation.}. Scale invariance is a desirable property of RFI detection algorithms, because (a) it implies the method can be applied on data with different resolutions without changing parameters and (b) the time and frequency scale of RFI itself can be arbitrary, so any method to detect RFI should work equally well for RFI at different scales. In practice, RFI seems to behave in a more or less scale-invariant manner at the resolution of LOFAR, as for example can be seen in Fig.~\ref{fig:wsrt-example}, so we should also use a scale-invariant method to detect it. This scale-invariant behaviour of RFI breaks down at high time and frequency resolutions, at which many features become diagonal in the time-frequency plane.

The proposed technique was first mentioned in \citet{LOFAR-RFI-pipeline}, as it is part of the AOFlagger, which is the default LOFAR RFI detection pipeline. In that article the operation was referred to as a dilation, however, it does not strictly adhere to all the properties of a morphological dilation. For example, we will see that the operator $\rho$ is not distributive over the union set operator: $\rho(X \cup Y) \neq \rho(X) \cup \rho(Y)$ for some $X$ and $Y$. Because a rank operator flags points for which the number of flagged points in a neighbourhood exceeds a threshold (\citealp[\textsection3.4]{mathematical-morphology-goutsias}, \citealp{morphological-operators-rank-filters}), we will refer to the operator $\rho$ as the scale-invariant rank (SIR) operator.

In this paper, we will describe the method in-depth and analyse its effectiveness. In \citet{LOFAR-RFI-pipeline}, it was mentioned that the full algorithm has a time complexity of $\mathcal{O}(N^2)$, $N$ being the input size of the SIR operator, but by making the algorithm less accurate, an implementation of $\mathcal{O}(N \times \log N)$ was mentioned to be possible. Here, we will introduce a faster algorithm with linear time complexity, which is also an \emph{exact} implementation of the SIR operator.

\subsection{Description}
\begin{figure*}
 \begin{center}  
  \subfloat[]{ \label{fig:gaussian-broadband-no-noise}
   \includegraphics[height=45mm]{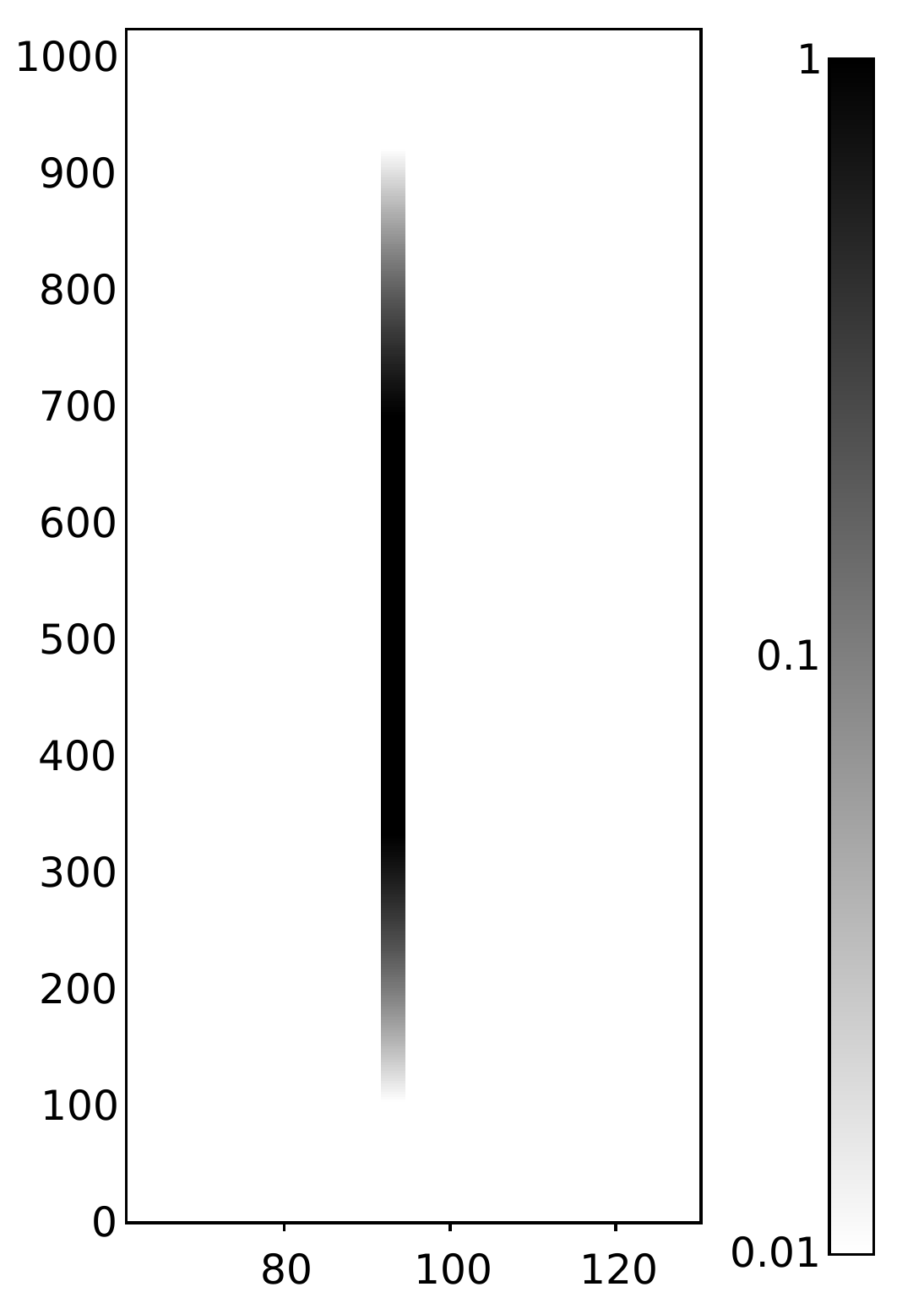}
  }\hspace{3mm}\nolinebreak%
  \subfloat[]{ \label{fig:gaussian-broadband-with-noise}
   \includegraphics[height=45mm]{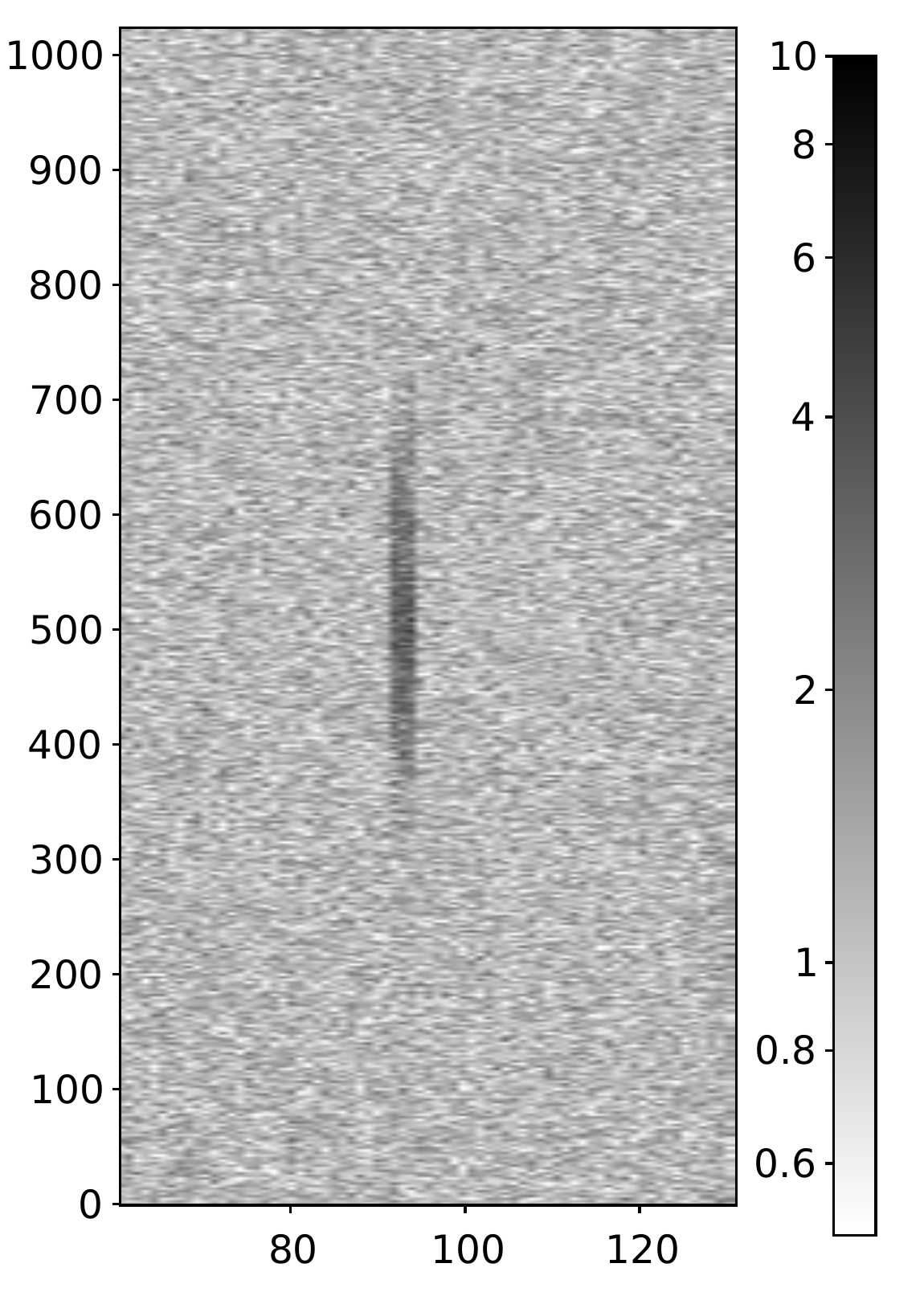}
   }\hspace{3mm}\nolinebreak%
  \subfloat[]{ \label{fig:gaussian-broadband-flagged}
   \includegraphics[height=45mm]{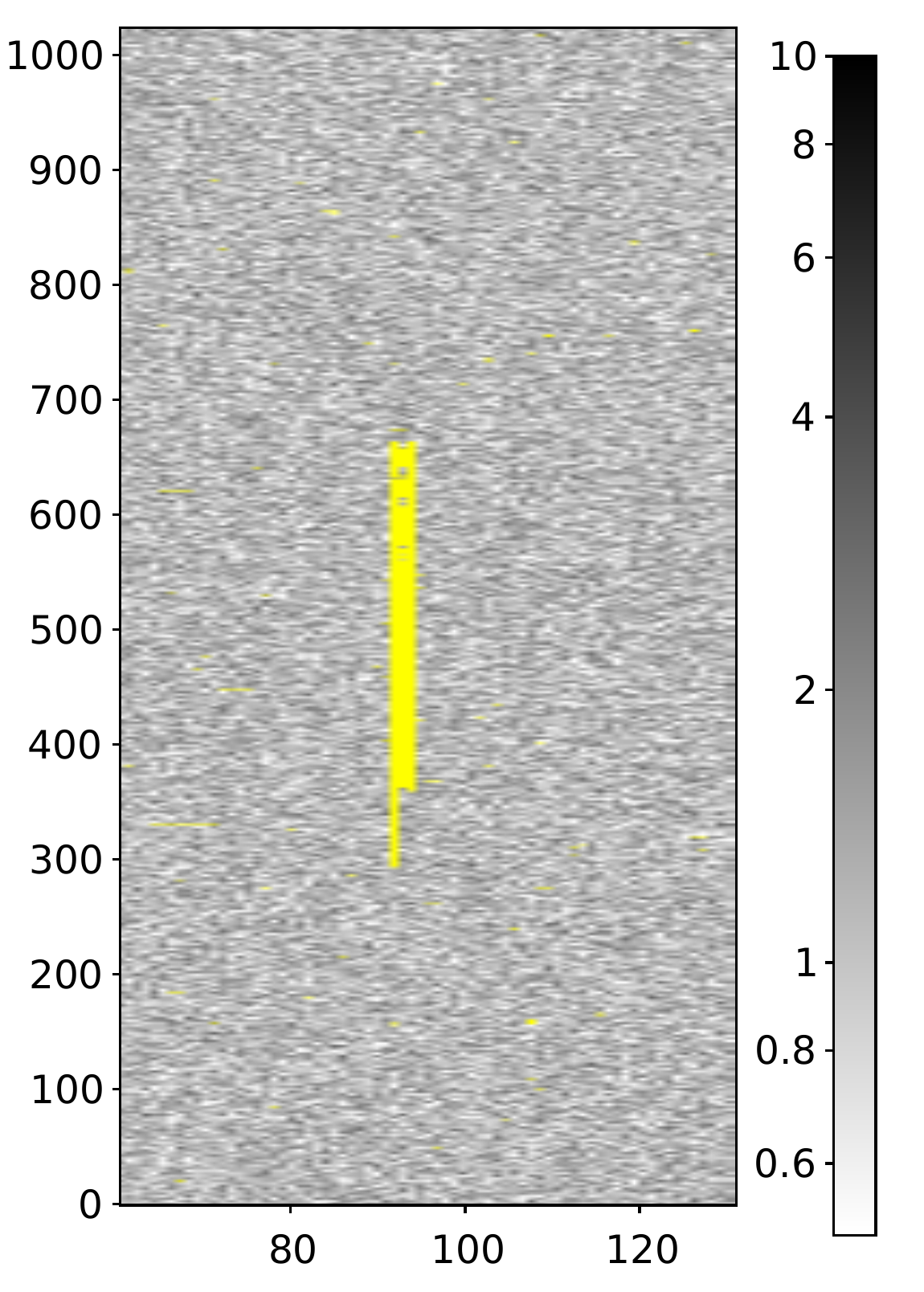}
   }\hspace{3mm}\nolinebreak%
  \subfloat[]{ \label{fig:gaussian-broadband-dilated}
   \includegraphics[height=45mm]{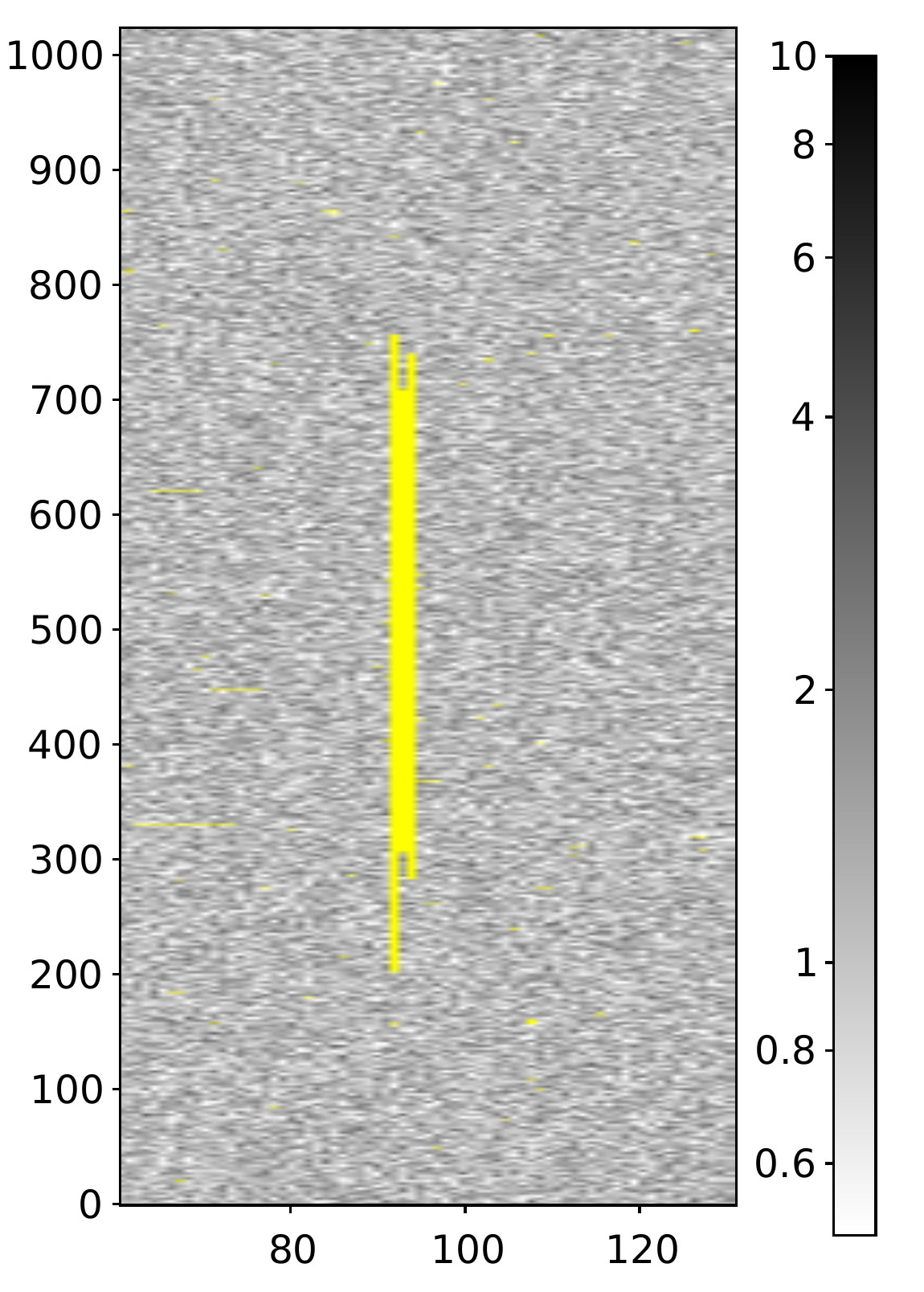}
   }
\end{center}
 \vspace{-0.5cm}
 \caption[]{Simulation of a typical broadband RFI feature with Gaussian frequency profile as used in the ROC analysis. Panel (a): isolated RFI feature; panel (b): when noise is added, a part of the feature becomes undetectable; panel (c): flagged with the \texttt{SumThreshold} method; panel (d): with SIR operator applied, parameter $\eta=0.2$. }
 \label{fig:gaussian-broadband}
\end{figure*}

Consider $F$, a set of positions in the time-frequency domain, such that a sample at time $t$ and frequency $\nu$ has been flagged when $(t,\nu) \in F$. Assume $F$ is the result of a statistical detection algorithm, such as the \texttt{SumThreshold} algorithm. We will apply the SIR operator in time and frequency directions separately, and define the sets $\Theta_t$ and $\Phi_\nu$ to contain the flags of a slice in time and frequency direction:
\begin{align}
\Theta_t \equiv & \left\{ (s,\nu) \in F \mid s=t \right\}, \\
\Phi_\nu \equiv & \left\{ (t,\mu) \in F \mid \mu = \nu \right\}.
\end{align}
A single one-dimensional set $\Theta_t$ or $\Phi_{\nu}$ is the input for the SIR operator. The operator considers a sample to be contaminated with RFI when the sample is in a subsequence of mostly flagged samples. To be more precise, it will flag a subsequence when more than $(1-\eta) N$ of its samples are flagged, with $N$ the number of samples in the subsequence and $\eta$ a constant, $0 \le \eta \le 1$. Using $\rho$ to denote the operator, the output $\rho(X)$ can be formally defined as
\begin{eqnarray} \label{eq:scale-invariant-rank-operator}
\rho(X)&\equiv&\bigcup\bigg\{[Y1,Y2)\mid \\ \notag
& & \#\left(X\cap[Y1,Y2)\right)\ge(1-\eta)(Y2-Y1)\bigg\},
\end{eqnarray}
with $[Y_1, Y_2)$ a half-open interval of $\Theta_t$ or $\Phi_\nu$, and the hash symbol $\#$ denoting the count-operator that returns the number of elements in the set. In words, Equation~\ref{eq:scale-invariant-rank-operator} defines $\rho(X)$ to consist of all the samples that are in an interval $[Y1,Y2)$, in which the ratio of samples in the input $X$ is greater or equal than $(1-\eta)$. Parameter $\eta$ represents the aggressiveness of the method: with $\eta=0$, no additional samples are flagged and $\rho(X) = X$. On the other hand, $\eta=1$ implies all samples will be flagged. Figure~\ref{fig:gaussian-broadband} shows an example of a simulated Gaussian broadband RFI feature, and the input and output of the SIR-operator.

\begin{figure}
 \begin{center}  
  \includegraphics[width=7cm]{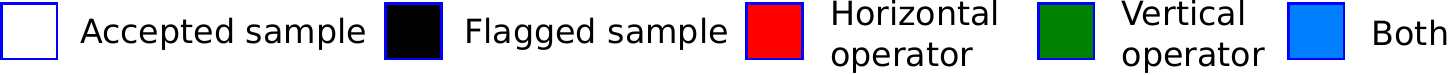}\\%
  \subfloat[Input]{ \label{fig:transfer-original}
   \includegraphics[width=1.5cm]{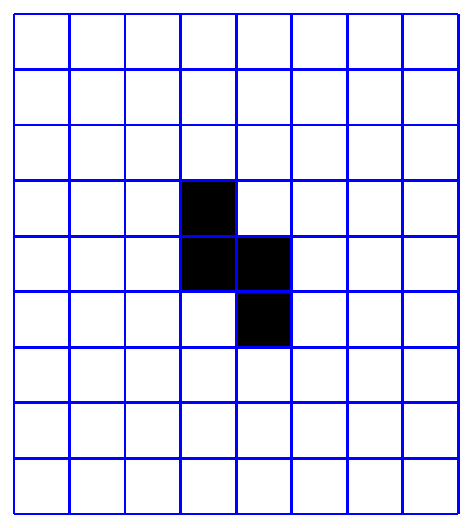}\hspace{0.5cm}
  }%
  \subfloat[Union]{ \label{fig:transfer-union}
   \includegraphics[width=1.5cm]{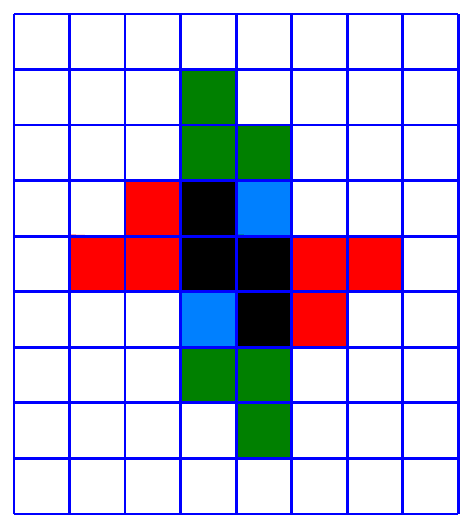}\hspace{0.5cm}
   }\\%
  \subfloat[Horizontal\newline first]{ \label{fig:transfer-first-horizontally}
   \includegraphics[width=1.5cm]{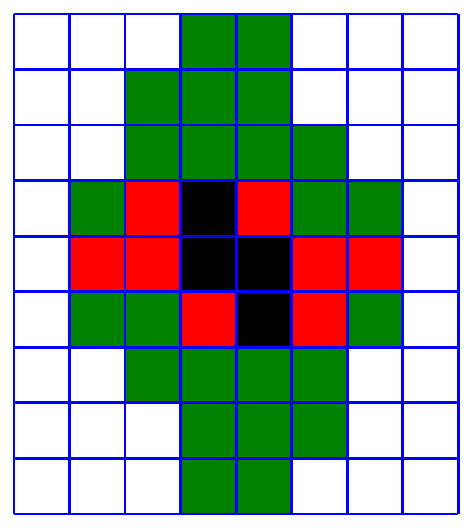}\hspace{0.5cm}
   }%
  \subfloat[Vertical first]{ \label{fig:transfer-first-vertically}
   \includegraphics[width=1.5cm]{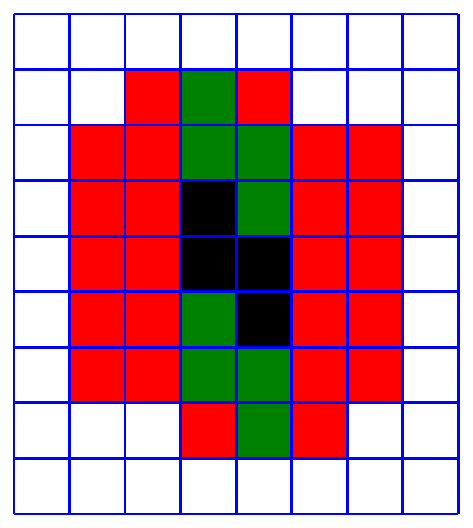}\hspace{0.5cm}
   }%
 \end{center}
 \caption[]{Example outputs of the SIR operator in which the one-dimensional output has been combined in three different ways. Panel~\subref{fig:transfer-original} is the input, panel~\subref{fig:transfer-union} shows the result of performing a union on the outputs of both directions, and in panels~\subref{fig:transfer-first-horizontally} and \subref{fig:transfer-first-vertically}, the SIR operator was first applied in, respectively, the horizontal and vertical direction. Parameter $\eta$ was $0.5$ in this example. }
 \label{fig:twodimtransfer}
\end{figure}

\begin{figure*}
 \begin{center}  
  \subfloat[Original data]{ \label{fig:lofar-rank-example-2dim-original}
   \includegraphics[height=35mm]{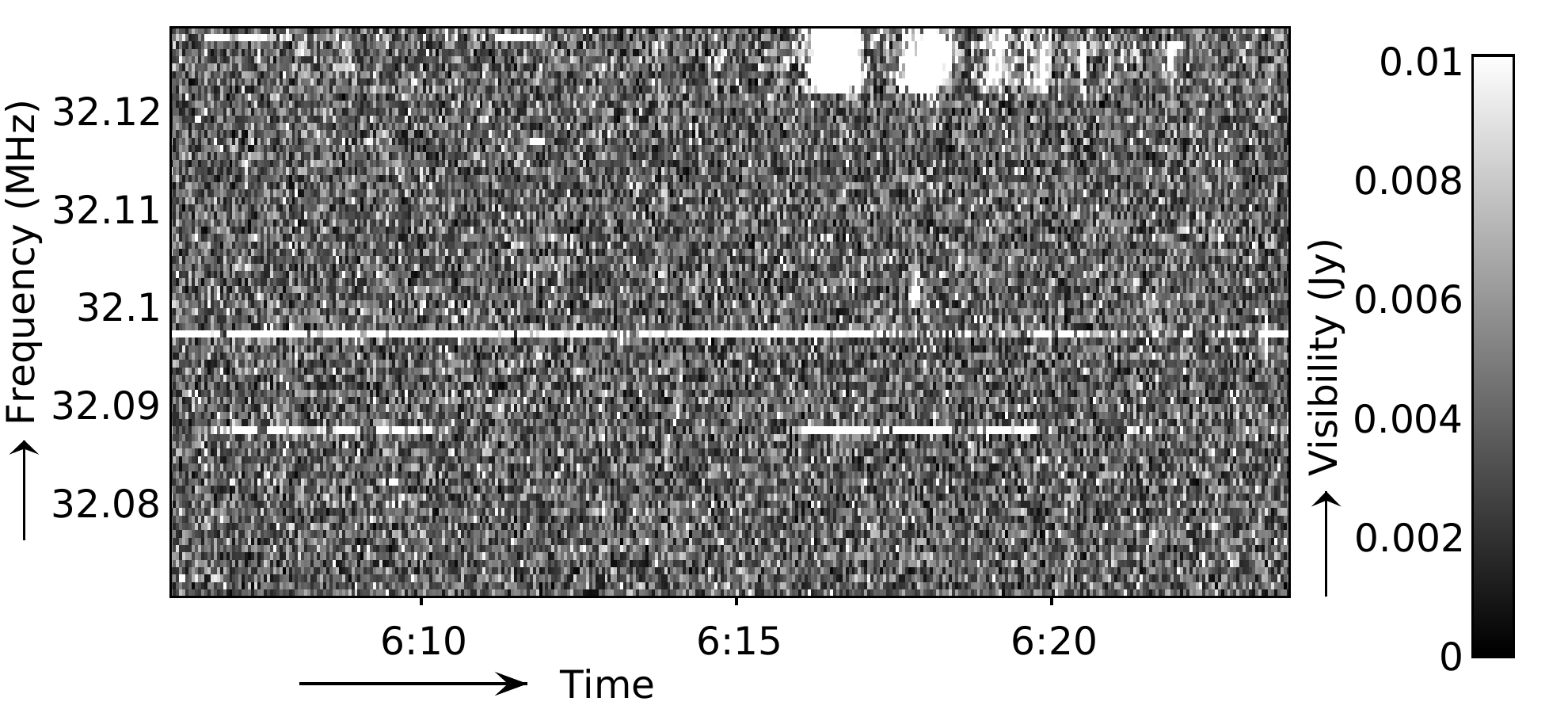}
  }\hspace{3mm}\nolinebreak%
  \subfloat[Intersection]{ \label{fig:lofar-rank-example-2dim-intersection}
   \includegraphics[height=35mm]{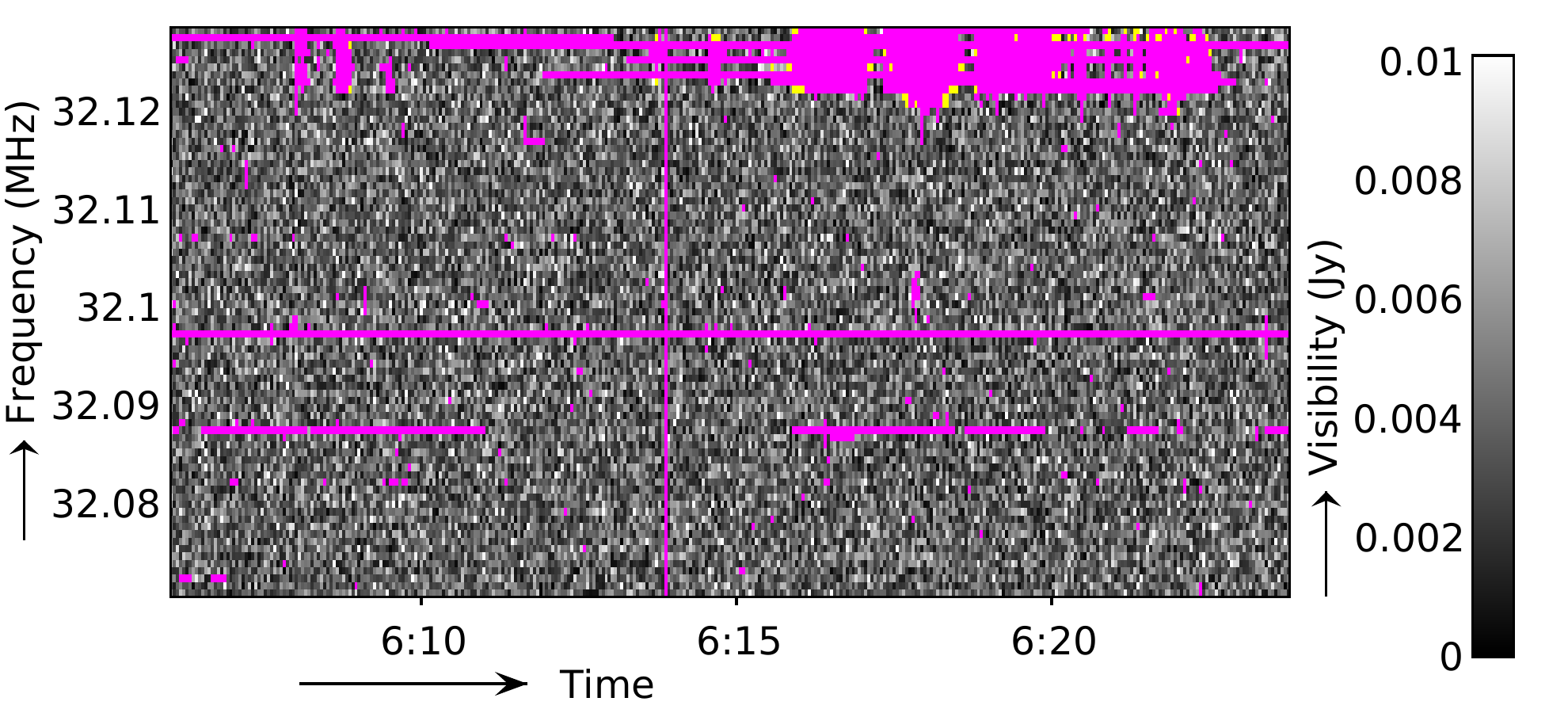}
   }\vspace{2mm}\\%
  \subfloat[Union]{ \label{fig:lofar-rank-example-2dim-union}
   \includegraphics[height=35mm]{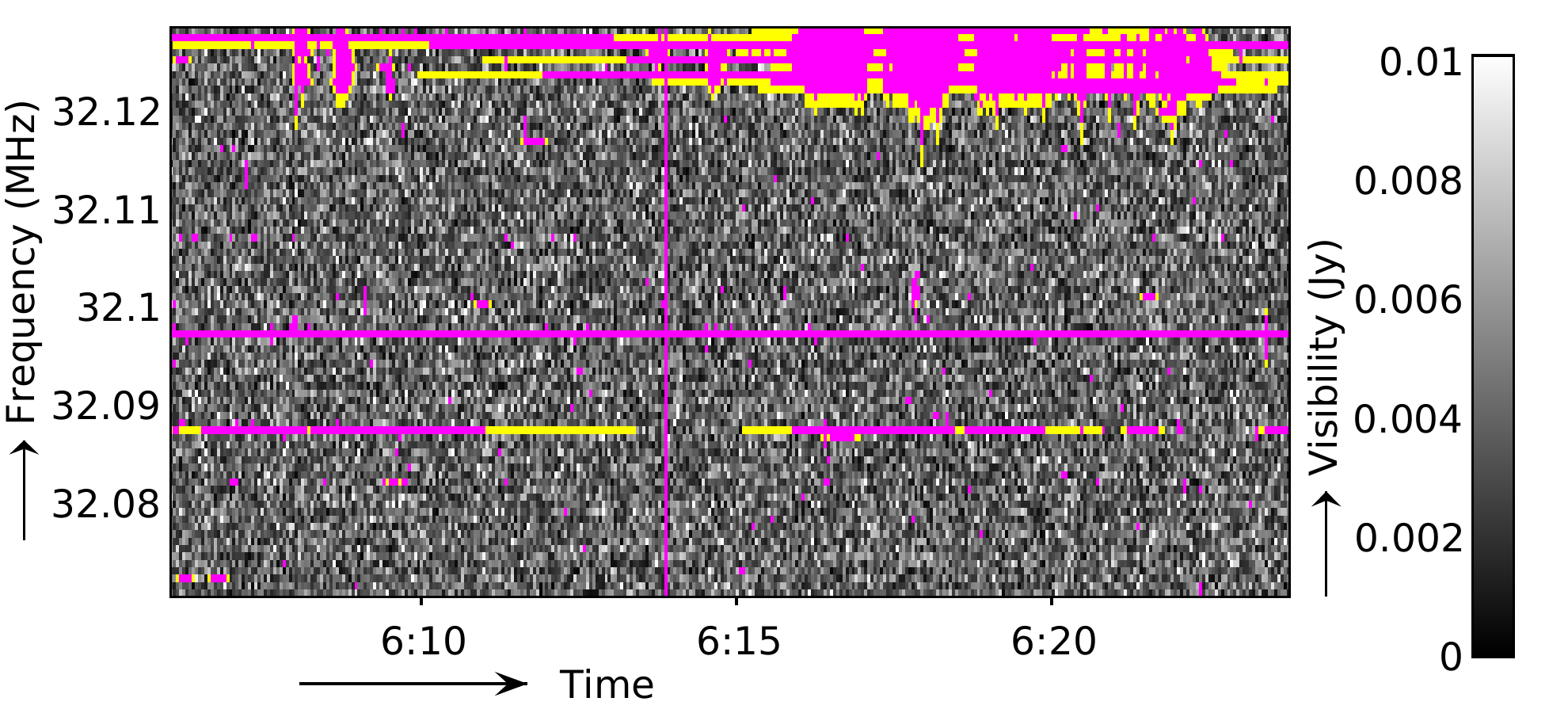}
   }\hspace{3mm}\nolinebreak%
  \subfloat[Time first]{ \label{fig:lofar-rank-example-2dim-time-first}
   \includegraphics[height=35mm]{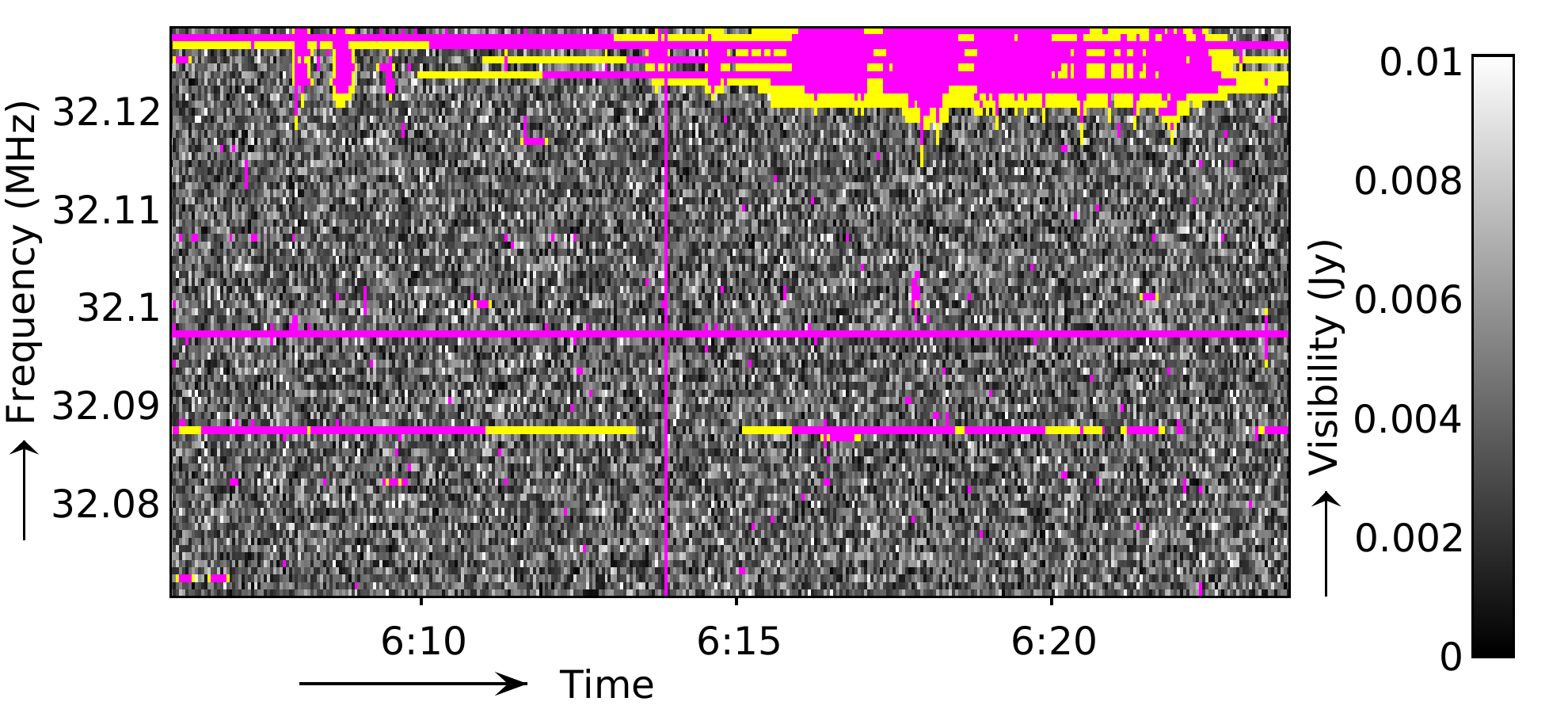}
   }\vspace{2mm}\\%
  \subfloat[Frequency first]{ \label{fig:lofar-rank-example-2dim-frequency-first}
   \includegraphics[height=35mm]{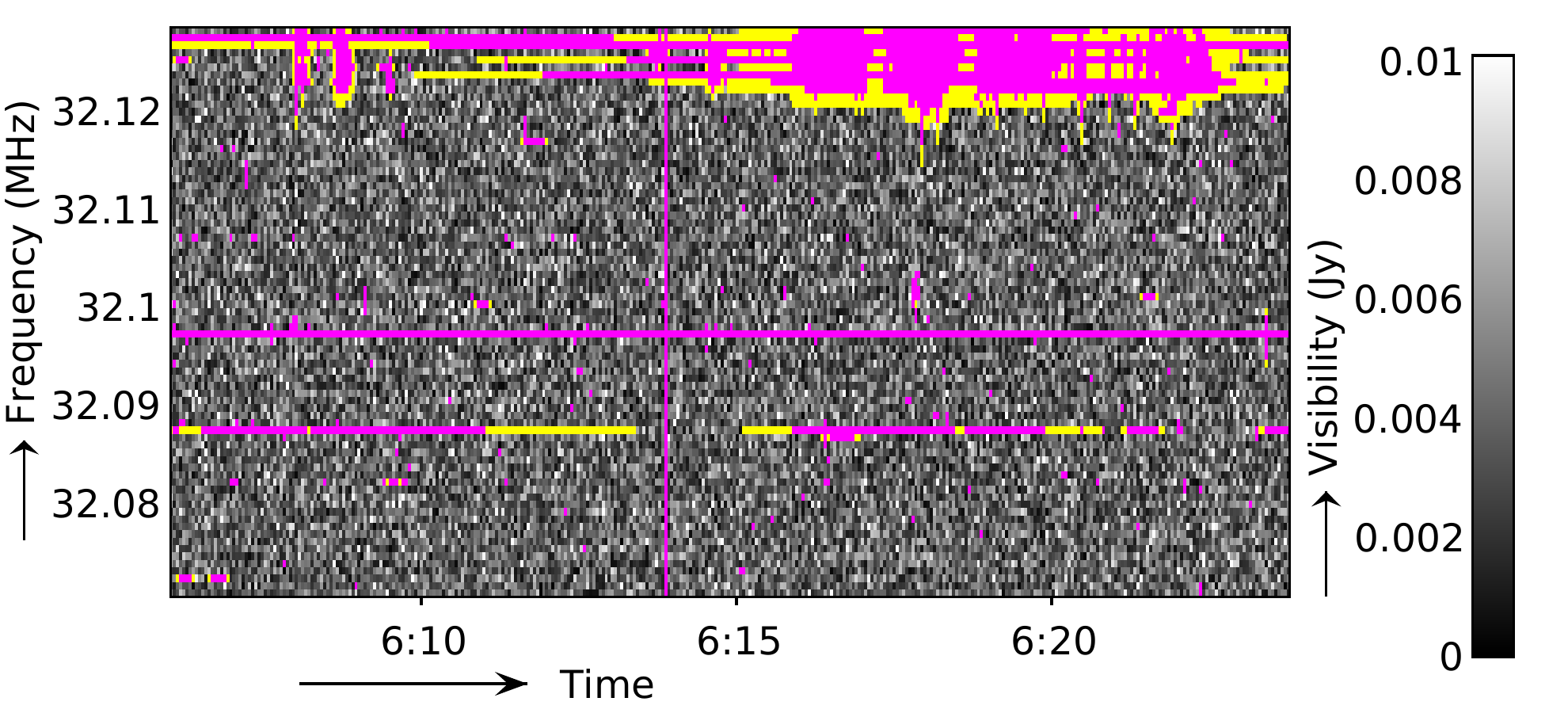}
   }\hspace{3mm}\nolinebreak
  \subfloat[Union of (d) and (e)]{ \label{fig:lofar-rank-example-2dim-union-of-sequential}
   \includegraphics[height=35mm]{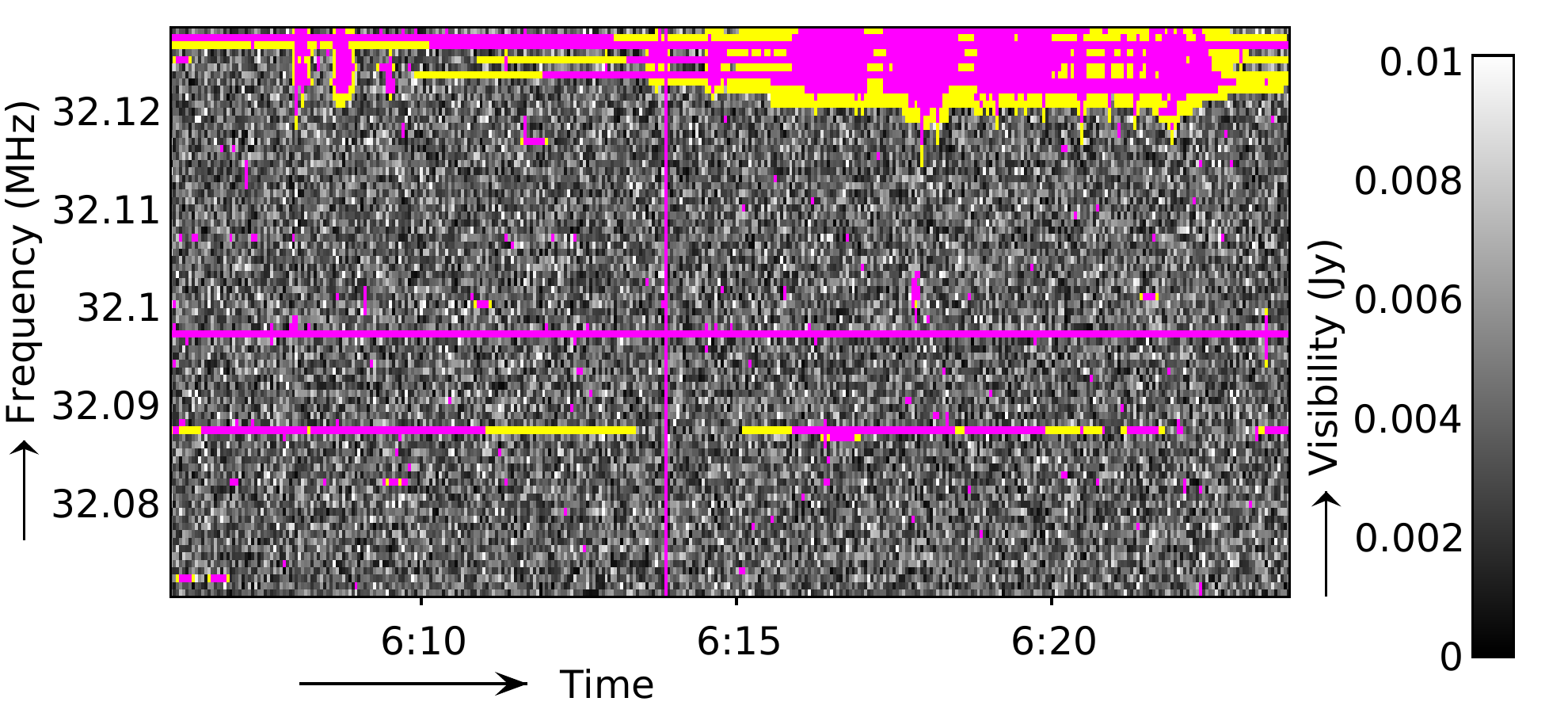}
   }
\end{center}
 \vspace{-0.5cm}
 \caption[]{Example of the SIR operator applied on a LOFAR observation, displaying five different methods to make the SIR operator two-dimensional. The visibilities shown are from baseline CS003 $\times$ CS007 of a LOFAR low-band-antenna (LBA) observation with 3s $\times$ 0.8 kHz resolution. This observation part was selected as an example because it has a two-dimensional RFI structure. Such RFI is less common, hence this is not a typical case. With the exception of the intersection, there is no difference between the different methods on the thin lines below 32.1 MHz. Applying the operator sequentially (panels d, e and f) is more aggressive for the two-dimensional structures, as it will flag samples that have diagonal neighbours that are flagged. Intersecting the two methods (panel b) will only flag concave samples. Pink is pre-flagged by the \texttt{SumThreshold} method, yellow is added by the SIR operator. A value of $\eta=0.2$ was used in this example.}
 \label{fig:lofar-rank-example-2dim}
\end{figure*}

The one-dimensional outputs can be remapped to the original two-dimensional domain in various ways. A simple and useful way is to perform a logical union of $\Theta'_t=\rho(\Theta_t)$ and $\Phi'_\nu=\rho(\Phi_\nu)$, the flags on respectively the time and frequency outputs:
\begin{equation}
F' = \left( \bigcup\limits_{t} \Theta_t' \right) \cup \left( \bigcup\limits_{\nu} \Phi_\nu' \right).
\end{equation}

An alternative is to initially apply the SIR operator only in one direction, i.e., on the sets that correspond with either the time or frequency direction, and subsequently applying the SIR operator on the outputs of the first in the other direction. The latter is more aggressive than the former. The result also depends on which direction is processed first. The difference is demonstrated in Fig.~\ref{fig:twodimtransfer}, and an example of how that would work out on actual data is given in Fig.~\ref{fig:lofar-rank-example-2dim}. Optionally, the operator can be applied in frequency and time directions with different $\eta$, if one suspects that RFI acts differently in either direction.

\subsection{Properties \& parameters}
Consider the case in Equation~\eqref{eq:scale-invariant-rank-operator} when a subsequence of arbitrary length is flagged. Since the fraction of flagged samples within the subsequence is explicitly used to define its output, the operator is \emph{scale invariant}. Formally, an operator $\rho$ is scale invariant if and only if $\rho(\lambda X) = \lambda\rho(X)$, i.e., scaling the input $X$ with $\lambda$ followed by $\rho$ is equal to scaling the output $\rho(X)$ with $\lambda$. We will now give a formal proof of the scale invariance of the SIR operator.
\begin{proof}
With $\rho$ the SIR operator, we will scale the input $X$ with factor $\lambda$. If $\lambda=0$ we trivially have that $\rho(\lambda X)=\lambda\rho(X)$. Also, if $\rho(\lambda X)=\lambda\rho(X)$ for $\lambda>0$, it is not difficult to see that we also have $\rho(-\lambda X)=-\lambda\rho(X)$, as mirroring the input will mirror the output. Therefore, assume without loss of generality that $\lambda>0$. Now, substituting $X$ with $\lambda X$ in Equation~\eqref{eq:scale-invariant-rank-operator} results in
\begin{eqnarray} \notag
\rho(\lambda X)&=&\bigcup\left\{[Y1,Y2)\mid \right. \\ \notag
& & \left. \#\left(\lambda X\cap[Y1,Y2)\right)\ge(1-\eta)(Y2-Y1)\right\}.
\end{eqnarray}
By using $Z_1$=$Y_1/\lambda$ and $Z_2$=$Y_2/\lambda$, this can be rewritten to
\begin{eqnarray} \notag
\rho(\lambda X)&=&\bigcup\left\{[\lambda Z1, \lambda Z2)\mid \right. \\ \notag
& & \left. \#\left(\lambda X\cap[\lambda Z1,\lambda Z2)\right)\ge(1-\eta)(\lambda Z2 - \lambda Z1)\right\}.
\end{eqnarray}
If we assume continuous positions, both the left side and the right side of the comparison can be scaled by $1/\lambda$:
\begin{eqnarray} \notag
\rho(\lambda X)&=&\bigcup\left\{[\lambda Z1, \lambda Z2)\mid \right. \\ \notag
& & \left. \#\left(X\cap[Z1,Z2)\right)\ge(1-\eta)(Z2 - Z1)\right\},
\end{eqnarray}
and by using $[\lambda Z_1, \lambda Z_2)=\lambda[Z_1, Z_2)$ and the definition in Equation~\eqref{eq:scale-invariant-rank-operator}, this is equivalent to $\rho(\lambda X)=\lambda \rho(X)$.
\end{proof}
Because the time and frequency dimensions are obviously discrete and finite when applied on radio observations, in practice the scale invariance is limited by the resolution and size of the data.

The aggressiveness of the SIR operator can be controlled with the $\eta$ parameter, which can be chosen differently for the time and frequency directions. Because the method is scale invariant, the choice of $\eta$ can be made independent of the time and frequency resolutions of the input. The default $\eta$ parameter in the LOFAR pipeline is currently $\eta=0.2$ and is equal in both directions. This value has been determined by tweaking of the parameter and data inspection, e.g. by looking at the resulting time-frequency diagram and projections of the data variances. The results were checked for many observations. Higher values seem to remove too much data without much benefit, while some RFI is left undetected with lower values. The value works well for various telescopes and on different time and frequency resolutions. We will evaluate this setting in Section~\ref{sec:accuracy-analysis}.

Since most telescopes observe two linear or circular polarizations, RFI detection can consider each (correlated) polarization individually, and the operator can be applied on each produced mask independently. However, the flag masks are often kept equal between the polarizations, because calibration might become unstable when, for a particular sample, part of the polarizations are missing. Moreover, if one of the polarization feeds of the telescope has been affected by RFI, it is likely that the others also have been affected. For these reasons, the approach taken in the LOFAR pipeline is to use the \texttt{SumThreshold} method on all four cross-correlated polarizations (XX, XY, YX and YY) individually, then flag any sample for which at least one polarization has been flagged, and finally apply the SIR operator once on the combined mask. 

\subsection{The algorithm} \label{sec:algorithm}
A straightforward implementation of the operator in Equation~\eqref{eq:scale-invariant-rank-operator} is to test each possible contiguous subsequence. In this case, if $N$ is the number of samples in the sequence $\Theta_t$ or $\Phi_\nu$, $\mathcal{O}(N^2)$ sums of subsequences have to be tested. Since the sums of all subsets can also be constructed in quadratic time complexity, the total time complexity of a straightforward implementation is $\mathcal{O}(N^2)$. We will now show an algorithm that solves the problem with linear time complexity. The algorithm is somewhat similar to the maximum contiguous subsequence sum algorithm.
\pagebreak
\begin{center}
\textbf{Listing 1}: \textit{\small{Linear time complexity algorithm for the scale-invariant rank operator}}
\end{center}
\begin{small}
\texttt{\textbf{~~function} ScaleInvariantRankOperator}\vspace{1mm} \\ 
\texttt{
\begin{tabular}{lll}
~~&Input: \\
  &~~\(N\) &: Size \\
  &~~\(\Omega\) &: Input array of size \(N\) \\
  & & \hspace{1mm} (\(\Omega\)[\(i\)] = 1 \(\implies\) \(i\) is flagged,\\
  & & \hspace{3mm} \(\Omega\)[\(i\)] = 0 otherwise) \\
  &~~\(\eta\) &: Aggressiveness parameter \\
  &Output: \\
  &~~\(\Omega'\) &: Output flag array of size \(N\) \\
\end{tabular}}\begin{alltt}
 1:\textbf{begin}
     \textit{// Initialize \(\Psi\)}
 2:  \textbf{for} \(x=0\ldots{}N-1\) \textbf{do} \(\Psi\)[\(x\)] \(\leftarrow\) \(\eta\) - 1 + \(\Omega\)[\(x\)]

     \textit{// Construct an array \(M\) such that:}
     \textit{// \(M(x) = \sum j\in\{0\ldots{}x-1\}\): \(\Psi\)[\(j\)]}
 3:  \(M\)[\(0\)]\(\leftarrow0\)
 4:  \textbf{for} \(x=0\ldots{}N-1\) \textbf{do} \(M\)[\(x+1\)] \(\leftarrow\) \(M\)[\(x\)] + \(\Psi\)[\(x\)]

     \textit{// Construct array \(P\) such that:}
     \textit{// \(M\)[\(P\)[\(x\)]] \(= \min M\)[\(j\)]: \(0\le{}j\le{}x\)}
 5:  \(P\)[\(0\)]\(\leftarrow0\)
 6:  \textbf{for} \(x=1\ldots{}N-1\) \textbf{do}
 7:    \(P\)[\(x\)]\(\leftarrow{}P\)[\(x-1\)]
 8:    \textbf{if} \(M\)[\(P\)[\(x\)]] > \(M\)[\(x\)] \textbf{then} \(P\)[\(x\)]\(\leftarrow{}x\)
 9:  \textbf{end for}

     \textit{// Construct array \(Q\) such that:}
     \textit{// \(M\)[\(Q\)[\(x\)]] \(= \max M\)[\(j\)]: \(x<j<N\)}
10:  \(Q\)[\(N-1\)]\(\leftarrow{}N\)
11:  \textbf{for} \(x=N-2\ldots{}0\) \textbf{do}
12:    \(Q\)[\(x\)]\(\leftarrow{}Q\)[\(x+1\)]
13:    \textbf{if} \(M\)[\(Q\)[\(x\)]] < \(M\)[\(x+1\)] \textbf{then} \(Q\)[\(x\)]\(\leftarrow{}x+1\)
14:  \textbf{end for}

     \textit{// Flag sample \(x\) if \(M\)[\(Q\)[\(x\)]]\( - M\)[\(P\)[\(x\)]]\( \ge 0 \) }
15:  \textbf{for} \(x=0\ldots{}N-1\) \textbf{do}
16:    \textbf{if} \(M\)[\(Q\)[\(x\)]]-\(M\)[\(P\)[\(x\)]]\(\ge0\) \textbf{then}
17:      \(\Omega'\)[\(x\)]\(\leftarrow1\)
18:    \textbf{else}
19:      \(\Omega'\)[\(x\)]\(\leftarrow0\)
20:    \textbf{end if}
21:  \textbf{end for}

22:  \textbf{return} \(\Omega'\);
23:\textbf{end}
\end{alltt}
\end{small}

Listing 1 shows a direct algorithm to solve the SIR operator problem.

\begin{proof}
Using the definition of $\Omega(x)$ and $\Omega'(x)$, such that $1$ indicates that $x$ is flagged and $0$ that it is not, we can rewrite Equation~\eqref{eq:scale-invariant-rank-operator} as
\begin{eqnarray} \label{eq:rewritten-scale-invariant-rank-operator-with-omega}
\Omega'(x) =
\begin{cases}
1 & \mbox{ if } \exists Y_1\le x,Y_2 > x \textrm{, such that}\\
& \hspace{0.2cm} \sum\limits^{Y_2-1}_{y=Y_1} \Omega(y) \ge(1-\eta)(Y_2-Y_1) \\
0 & \mbox{otherwise.}
\end{cases}
\end{eqnarray}
In line 2, the array $\Psi(y)$ is initialized such that $\Psi(y)=\eta$ in case $y$ is flagged, and $\Psi(y)=\eta-1$ otherwise. Equation~\eqref{eq:rewritten-scale-invariant-rank-operator-with-omega} can now be rewritten to the following test:
\begin{equation} \label{eq:rewritten-scale-invariant-rank-operator}
\Omega'(x)=
\begin{cases}
1 & \mbox{if } \exists Y_1 \le x,~\exists Y_2 > x: \sum\limits_{y=Y_1}^{Y_2-1} \Psi(y) \ge 0 \\
0 & \mbox{otherwise.}
\end{cases}
\end{equation}
Line 3-4 initialize $M(x)$ for $0 \le x \le N$ to
\begin{equation}\notag
 M(x) = \sum_{j=0}^{x-1} \Psi(j),
\end{equation}
so that Equation~\eqref{eq:rewritten-scale-invariant-rank-operator} can be rewritten as
\begin{equation}
\Omega'(x)=
\begin{cases} \label{eq:rank-operator-with-m}
1 & \mbox{if } \exists Y_1 \le x,~\exists Y_2 > x:\\
& \hspace{0.5cm} M(Y_2)-M(Y_1) \ge 0 \\
0 & \mbox{otherwise.}
\end{cases}
\end{equation}
Because we are only interested in $\Omega'(x)$ in the range $0\le x < N$, we can limit the search for $Y_1$ and $Y_2$ to $0 \le Y_1 \le x < Y_2 \le N$.
There exists $Y_1$ and $Y_2$ in this range such that $M(Y_2)-M(Y_1)\ge0$, if and only if
\begin{equation}\notag
\max\limits_{y: x<y\le N} M(y) - \min\limits_{y:0\le y\le x} M(y) \ge 0.
\end{equation}
Lines 5-14 make sure that $P$ and $Q$ are initialized for $0\le x< N$, such that
\begin{eqnarray}\notag
P(x)&=&\underset{y\in 0\ldots x}{\operatorname{argmin}} \hspace{0.1cm} M(y),
\\ \notag
Q(x)&=&\underset{y\in x+1\ldots N}{\operatorname{argmax}} M(y).
\end{eqnarray}
Finally, this allows Equation~\eqref{eq:rank-operator-with-m} to be rewritten as
\begin{equation}
\Omega'(x)=
\begin{cases}
1 & \mbox{if } M(Q(x)) - M(P(x)) \ge 0 \\
0 & \mbox{otherwise,}
\end{cases}
\end{equation}
which is performed and returned in lines 15-23.
\end{proof}

The algorithm is $\Theta(N)$, and performs $3N$ additions or subtractions and $3N-2$ comparisons on floating point numbers. The algorithm uses the temporary arrays $\Psi$, $M$, $P$ and $Q$, each of size $N$, with the exception of $M$ which is of size $N+1$. Array $\Psi$ can be optimized away and the input $\Omega$ can be reused for output by assigning directly to it in lines 17 and 19. The total amount of temporary storage required is thus about $N$ floating point values and $2N$ index values, thus $\mathcal{O}(N)$. When the function is applied on a two-dimensional image, as in the case of RFI detection, the temporary storage is negligible, as the number of processed slices is usually much larger than one or two. If $\eta$ is expressed as a ratio of two integer values, it is possible to scale all values and only use integer math.

The algorithm has been implemented in C++ and takes around 40 lines of code\footnote{The implementation is part of the AOFlagger and can be downloaded from \url{http://www.astro.rug.nl/rfi-software}.}.

Because the problem is somewhat similar to the maximum contiguous subsequence sum \citep{programming-pearls-maximum-sum} and the all maximal contiguous subsequence sum problems, it might be possible to parallelize the algorithm by similar means, e.g. as in \citep{parallellized-maximal-subsequence-algorithm}. Moreover, parallel algorithms exist for the prefix sum/min/max calculations. For the specific application of RFI detection for LOFAR, the pipeline has already been maximally parallelized by flagging different baselines and/or sub-bands concurrently. Unlike parallelizing on the algorithm level, this requires no communication between the different processes.

\begin{figure}
 \begin{center}
 \includegraphics[width=8.7cm]{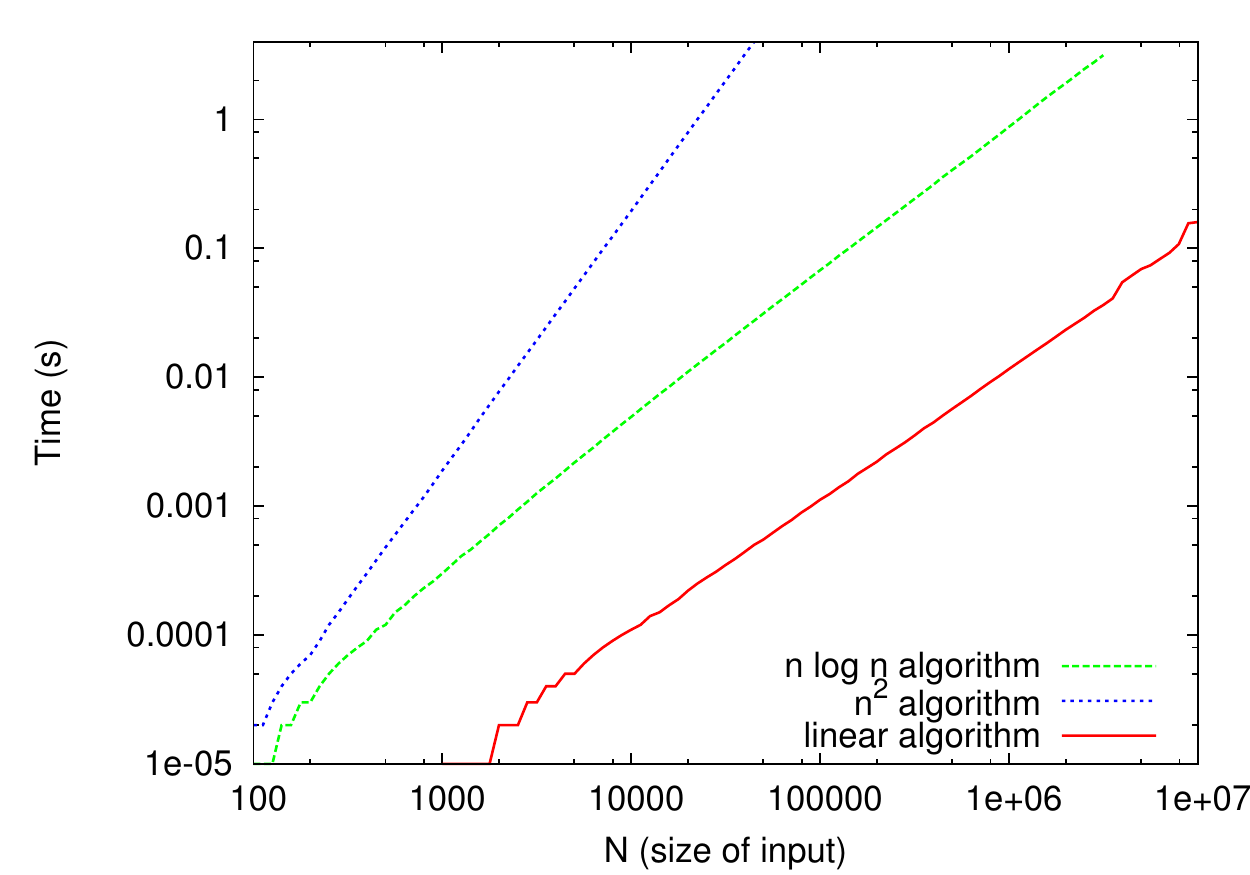}
 \end{center}
 \caption[]{Computation time versus input size with the different algorithms and fixed $\eta=0.2$. The average over 1000 runs was taken for each different configuration.}
 \label{fig:timing-plot}
\end{figure}

\section{Analysis \& results} \label{sec:results}
\subsection{Performance}
Figure~\ref{fig:timing-plot} displays the performance of the C++ implementations and compares the linear algorithm with the approximate $\mathcal{O}(N \log N)$ algorithm and the full quadratic algorithm. The measurements have been performed on a regular desktop with a 3.07 GHz Intel Core i7 CPU, using only one of its cores. The time complexities of the three algorithms for increasing $N$ behave as expected. The linear algorithm is faster in all cases, even for small $N$. The $\mathcal{O}(N \log N)$ time complexity algorithm is more than one order of magnitude slower at both small and large $N$. The linear algorithm has been executed with different values for $\eta$ and the results are shown in Fig.~\ref{fig:timing-plot}. Except for some slight variations
--- especially for $\eta=0$ --- the algorithm's speed is independent of $\eta$.

In the LOFAR pipeline, it takes 3.8 seconds to process a single sub-band for a single baseline, assuming 100,000 time steps and 256 channels (which is common). Of these 3.8 seconds, only 49 milliseconds (1.3\%) are spent applying the SIR operator. In common applications, an observation contains on the order of a 1,000 baselines and 250 sub-bands. The pipeline is heavily parallelized by concurrently flagging baselines over multiple cores and sub-bands over multiple cluster nodes. In this case, the pipeline's performance is dominated by disk access, and the relative contribution of the SIR operator is even smaller.

\subsection{Accuracy analysis} \label{sec:accuracy-analysis}
\begin{figure*}
 \begin{center}  
  \subfloat[]{ \label{fig:gaussian-broadband-feature}
   \includegraphics[height=45mm]{img/gaussian-broadband-with-noise}
   }\hspace{3mm}\nolinebreak%
   \subfloat[]{ \label{fig:sinusoidal-broadband-feature}
   \includegraphics[height=45mm]{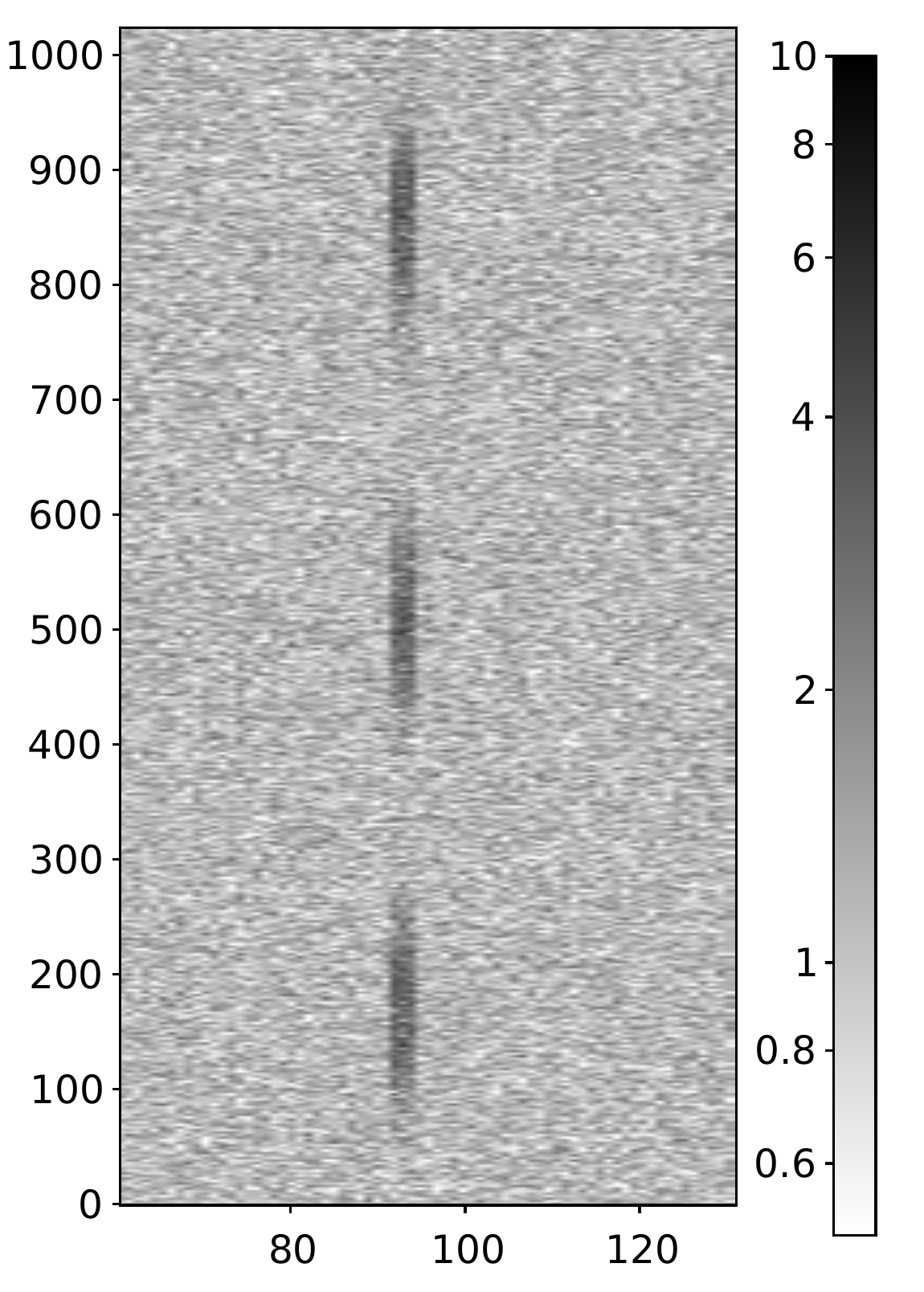}
   }\hspace{3mm}\nolinebreak%
   \subfloat[]{ \label{fig:slanted-feature}
   \includegraphics[height=45mm]{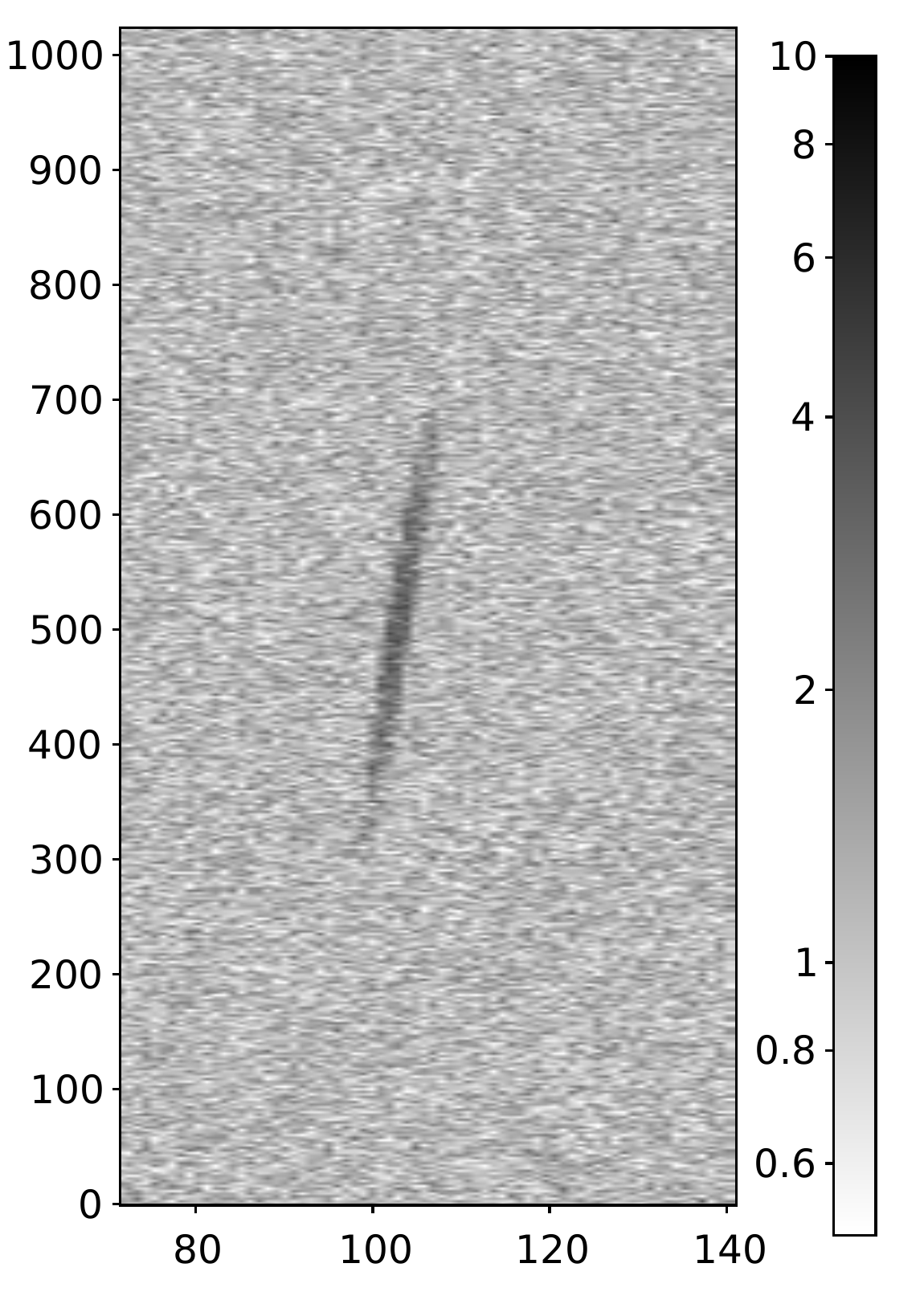}
   }\hspace{3mm}\nolinebreak%
   \subfloat[]{ \label{fig:burst-feature}
   \includegraphics[height=45mm]{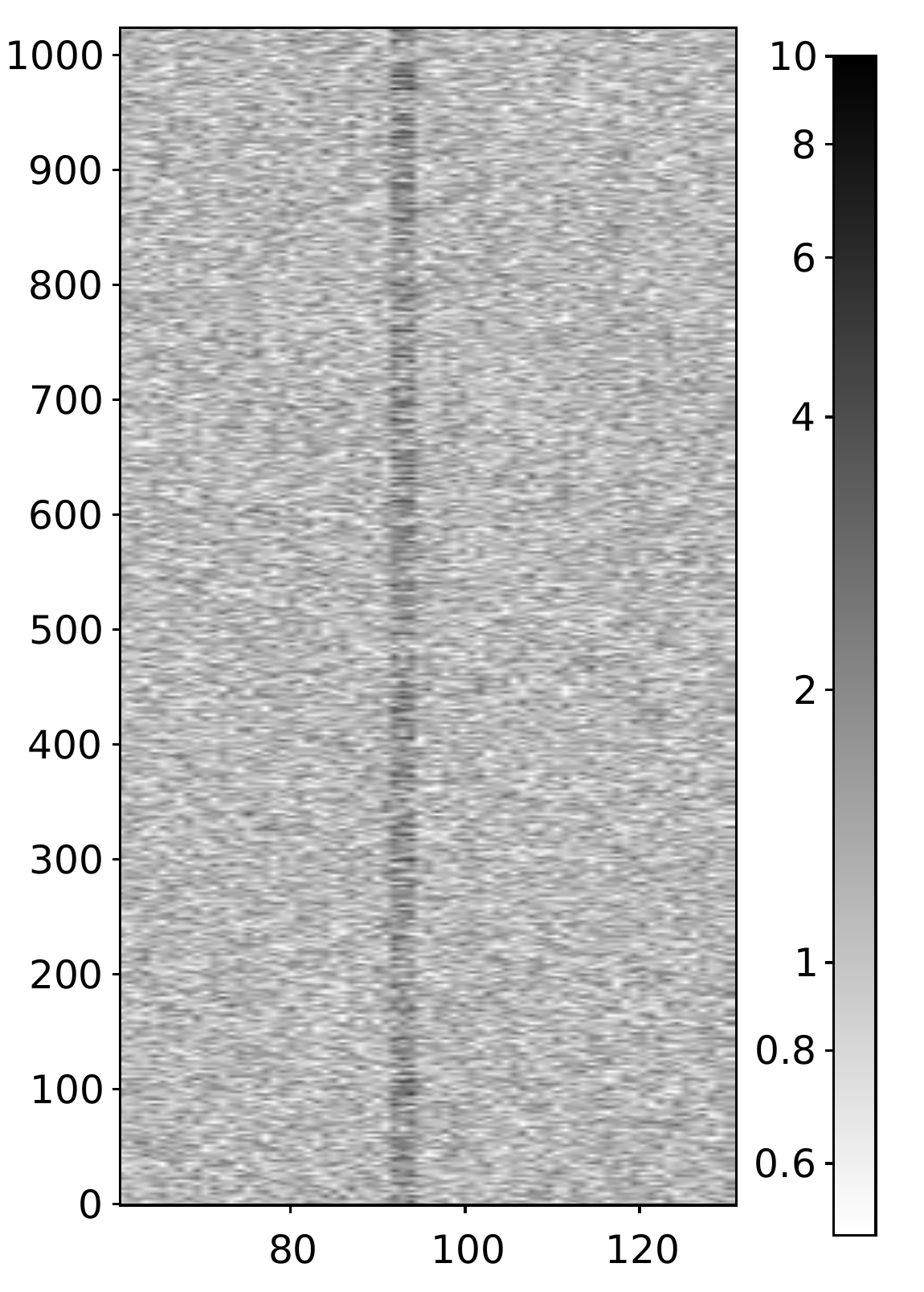}
   }
\end{center}
 \caption{Features used for the accuracy analysis. Panel (a): Feature with Gaussian slope; panel (b): Sinusoidal feature; panel (c): Slanted feature with Gaussian slope; panel (d): Burst feature with samples drawn from a Rayleigh distribution.}
 \label{fig:features}
\end{figure*}
\begin{figure*}
 \begin{center}
\subfloat[ROC curves. Solid lines: rank operator; dashed lines: dilation. ]{\label{fig:roc-analysis-roc}%
\includegraphics[width=120mm]{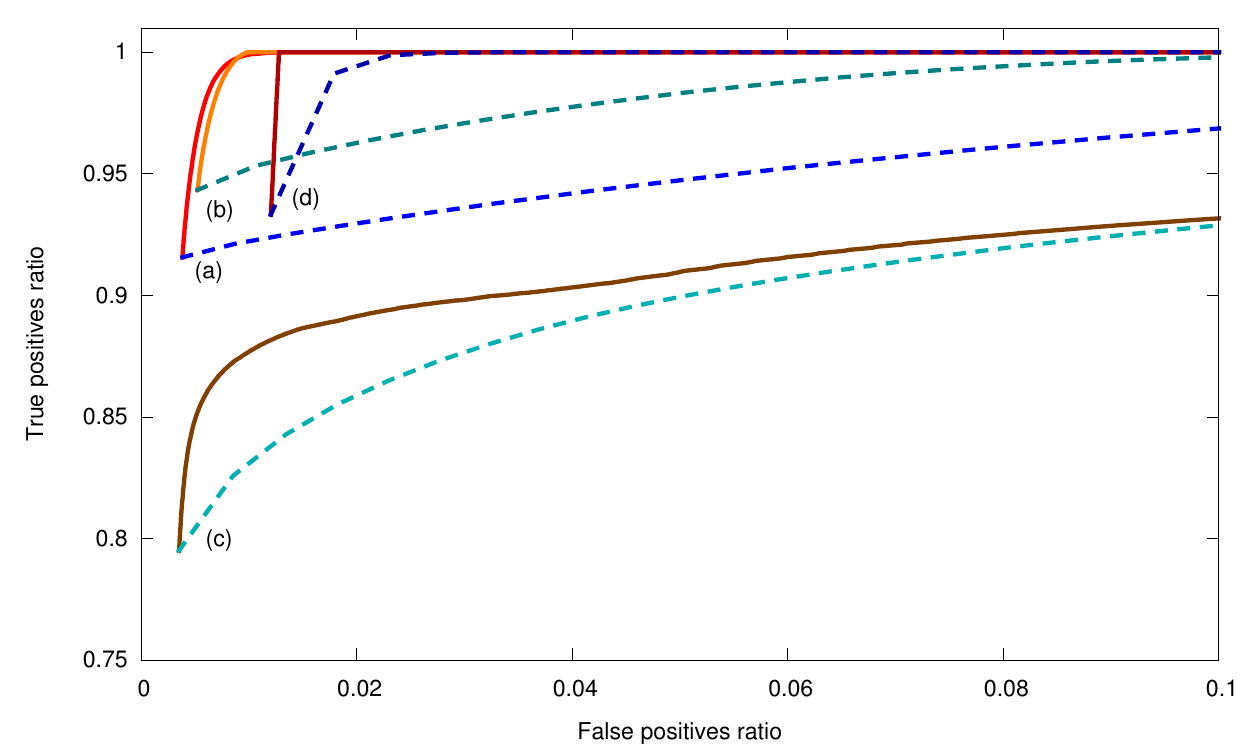}%
}\raisebox{2cm}{\includegraphics[width=35mm]{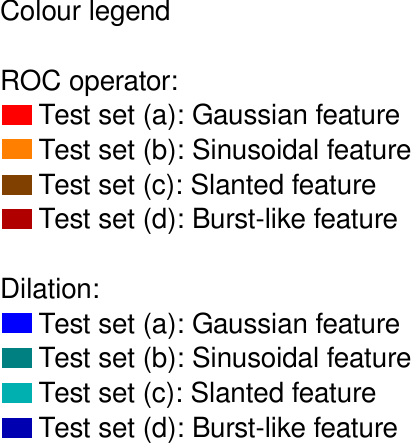}}\\
\subfloat[Influence of $\eta$ on the SIR operator. Solid lines: true positives (left axis); dashed lines: false positives (right axis).]{\label{fig:roc-analysis-rank-operator}%
\includegraphics[width=88mm]{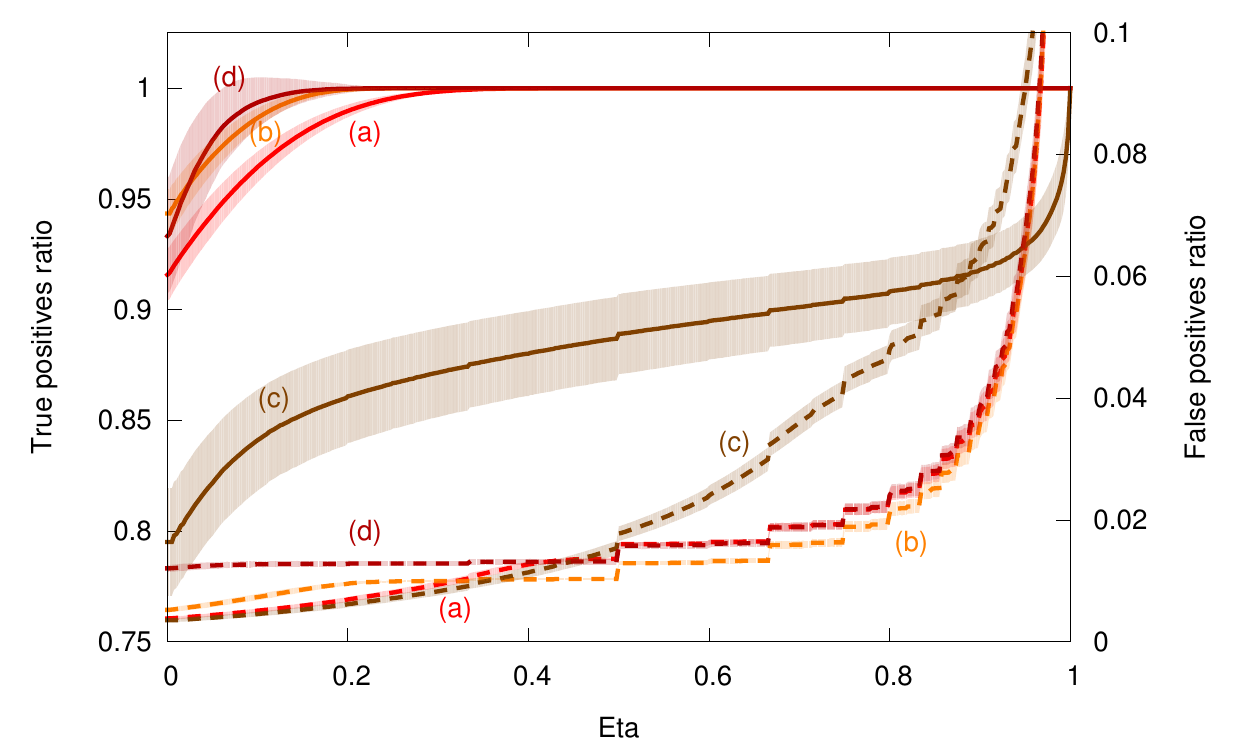}%
}\hspace*{5mm}%
\subfloat[Influence of kernel size on the dilation. Solid lines: true positives (left axis); dashed lines: false positives (right axis). ]{\label{fig:roc-analysis-dilation}%
\includegraphics[width=88mm]{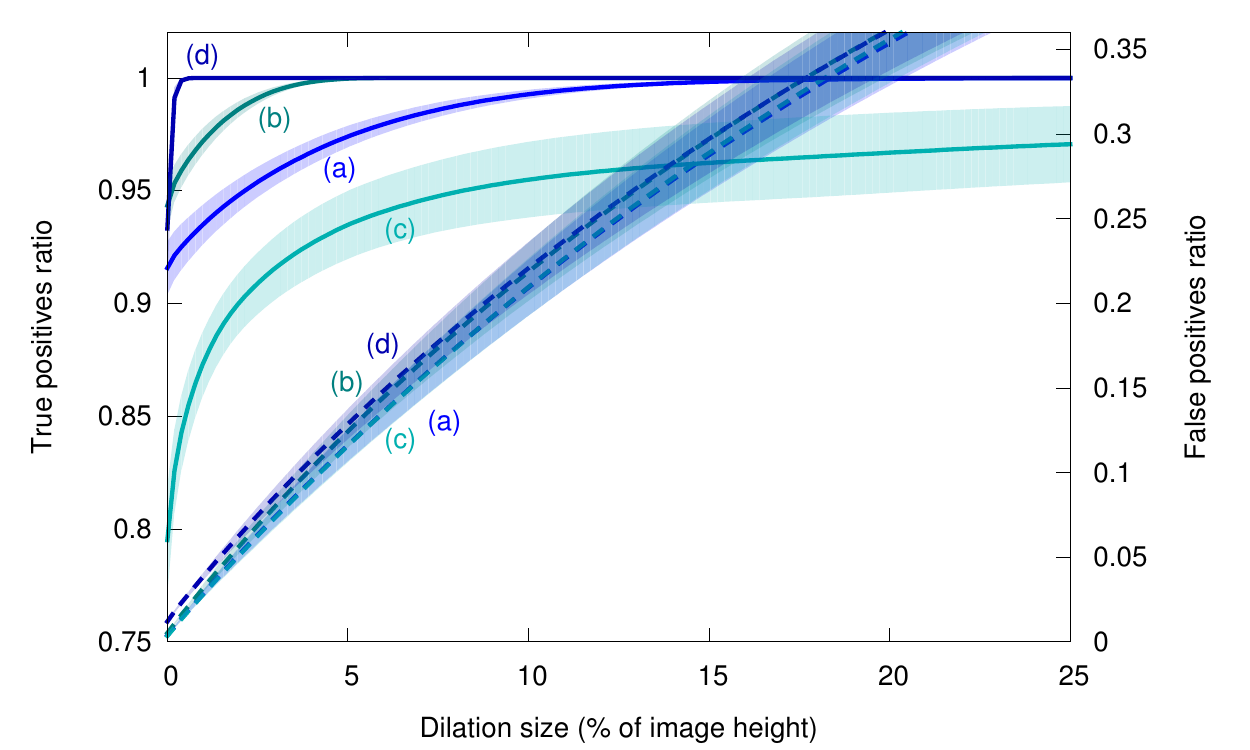}%
}
 \end{center}
 \caption[]{Analysis of the receiver operating characteristics of the SIR operator and a standard dilation on simulated data. Marks (a)--(d) correspond with the features shown in Fig.~\ref{fig:features}, respectively a Gaussian broadband feature, a sinusoidal feature, a slightly slanted Gaussian feature and a burst-like feature. The shadowed areas show $1\sigma$ levels over 100 runs.}
 \label{fig:roc-analysis}
\end{figure*}

\begin{figure*}
 \begin{center}  
  \subfloat[Horizontal SIR operator]{ \label{fig:wsrt-example-sir-horizontal}
  \includegraphics[width=55mm]{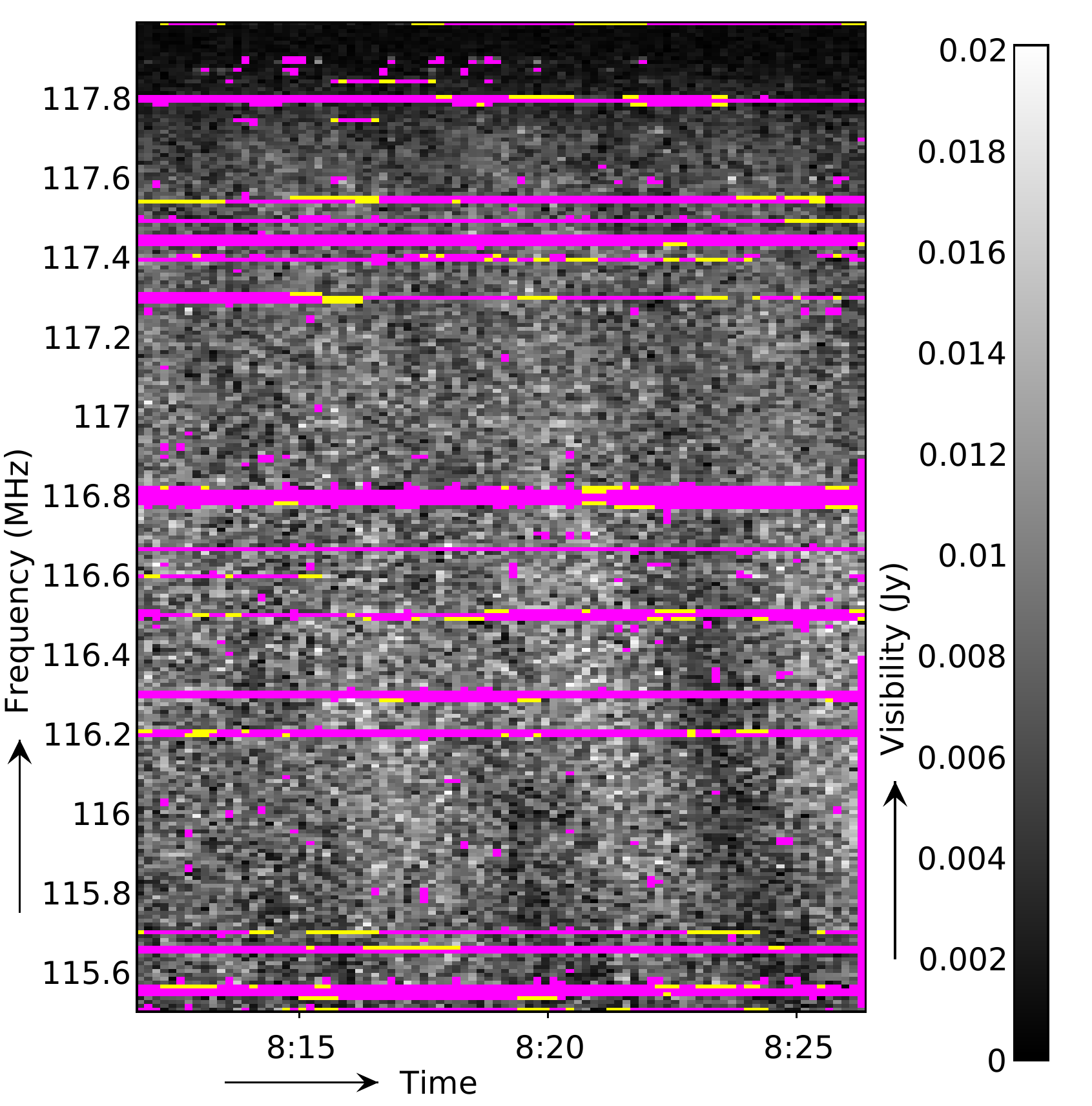}
  }\hspace{3mm}\nolinebreak%
  \subfloat[Vertical SIR operator]{ \label{fig:wsrt-example-sir-vertical}
  \includegraphics[width=55mm]{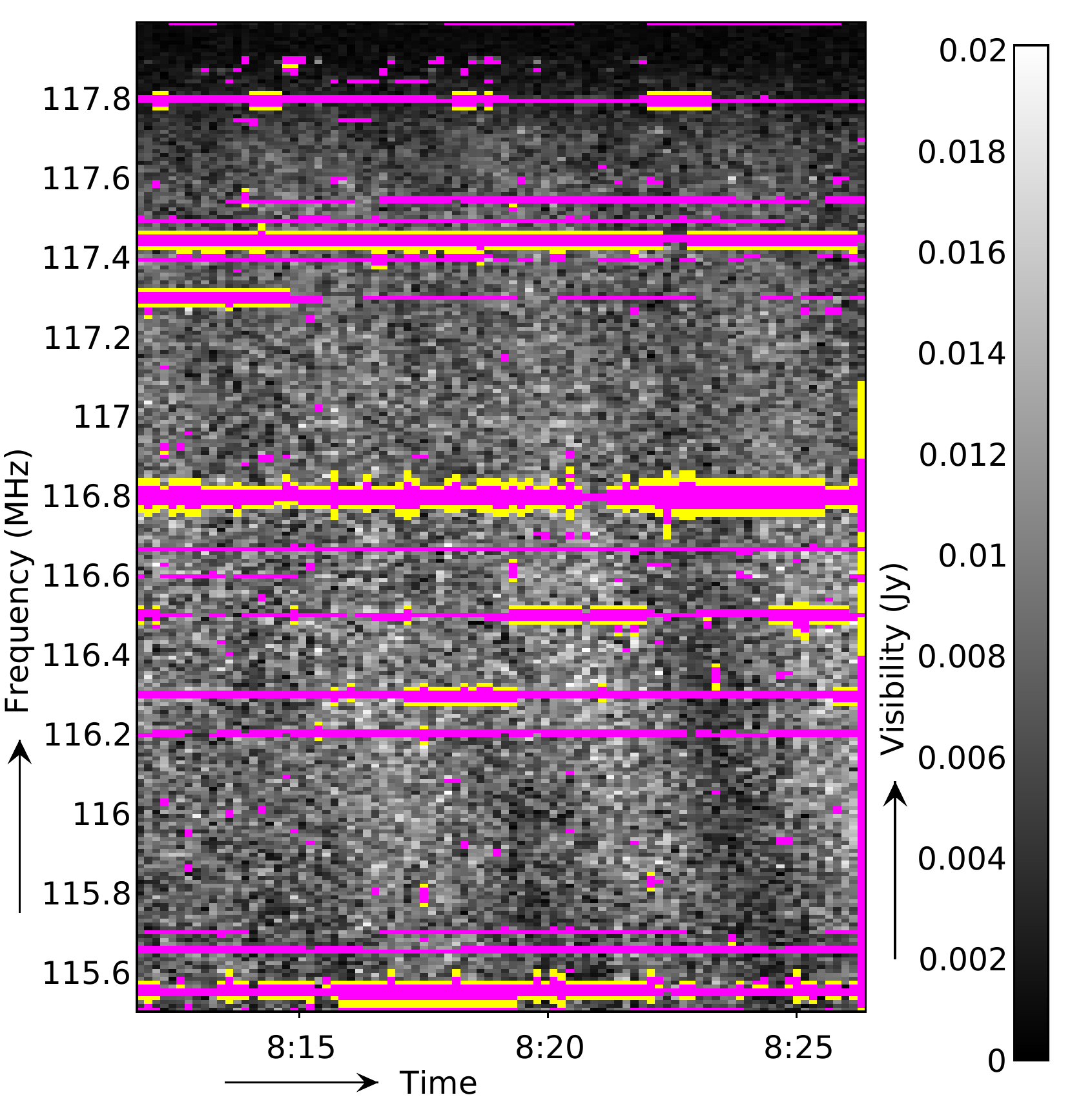}
  }\hspace{3mm}\nolinebreak%
  \subfloat[SIR operator in both directions]{ \label{fig:wsrt-example-sir-both}
  \includegraphics[width=55mm]{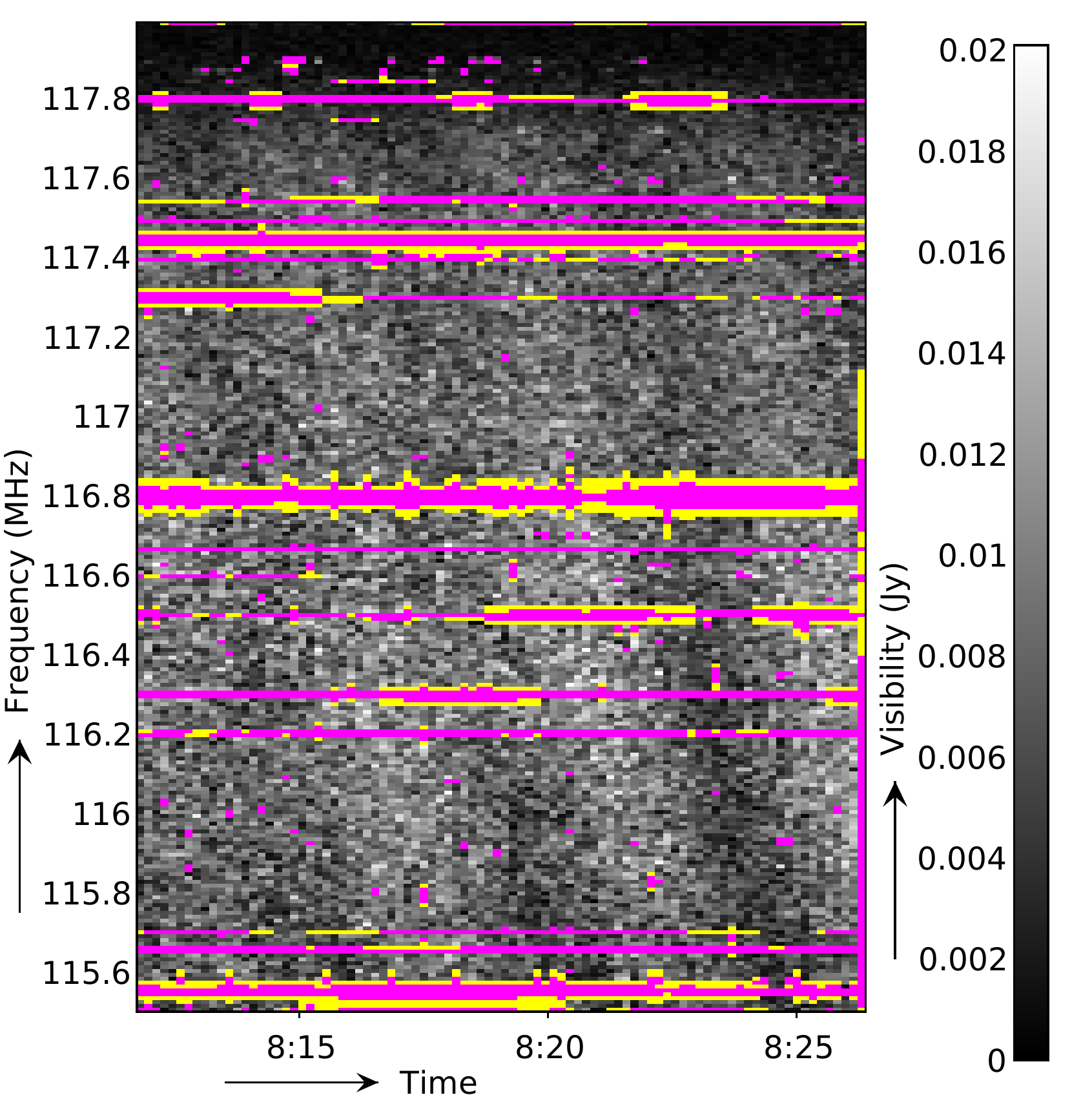}
  }\\%
  \subfloat[Horizontal dilation]{ \label{fig:wsrt-example-std-dilated-horizontal}
   \includegraphics[width=55mm]{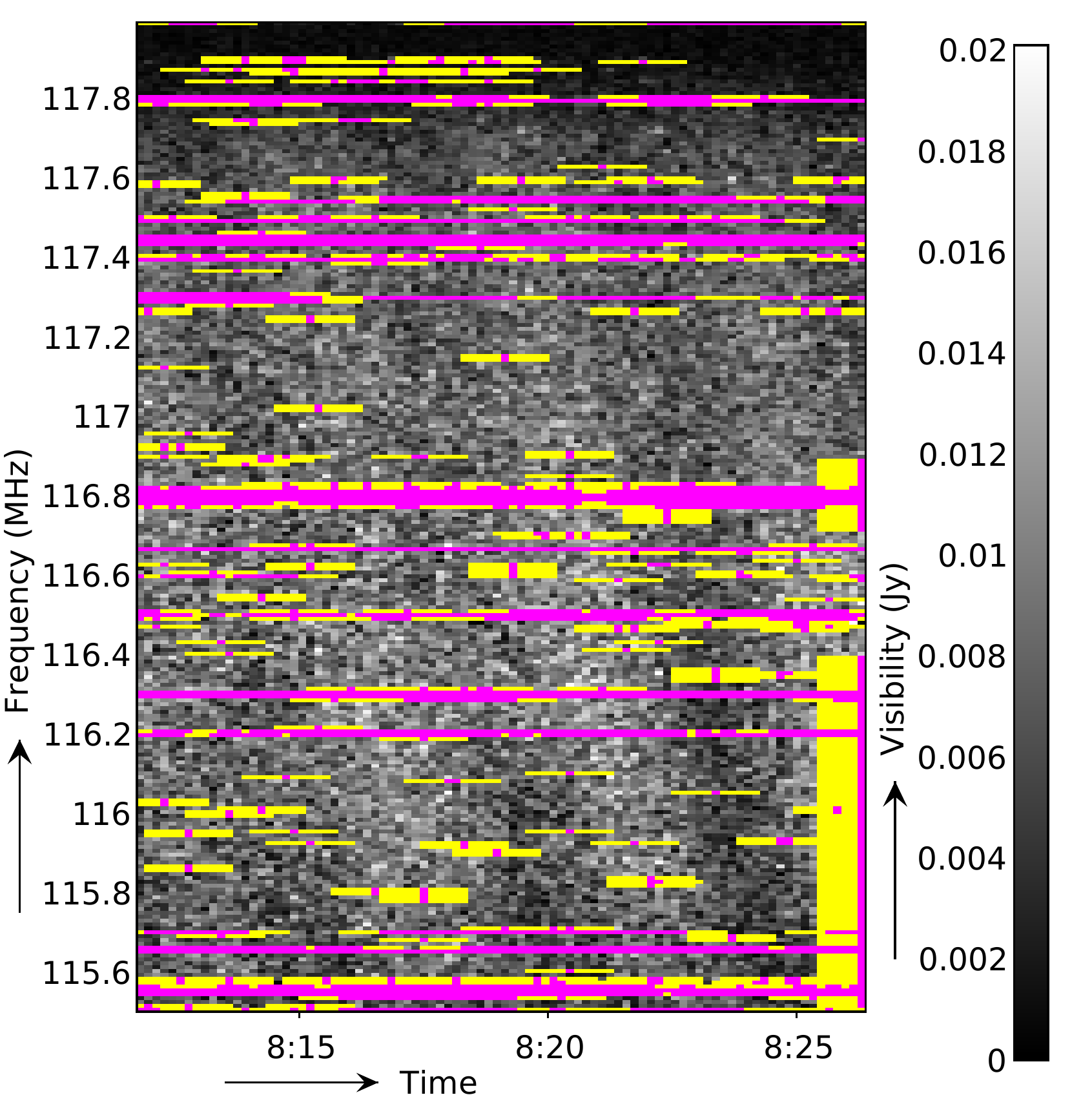}
  }\hspace{3mm}\nolinebreak%
  \subfloat[Vertical dilation]{ \label{fig:wsrt-example-std-dilated-vertical}
   \includegraphics[width=55mm]{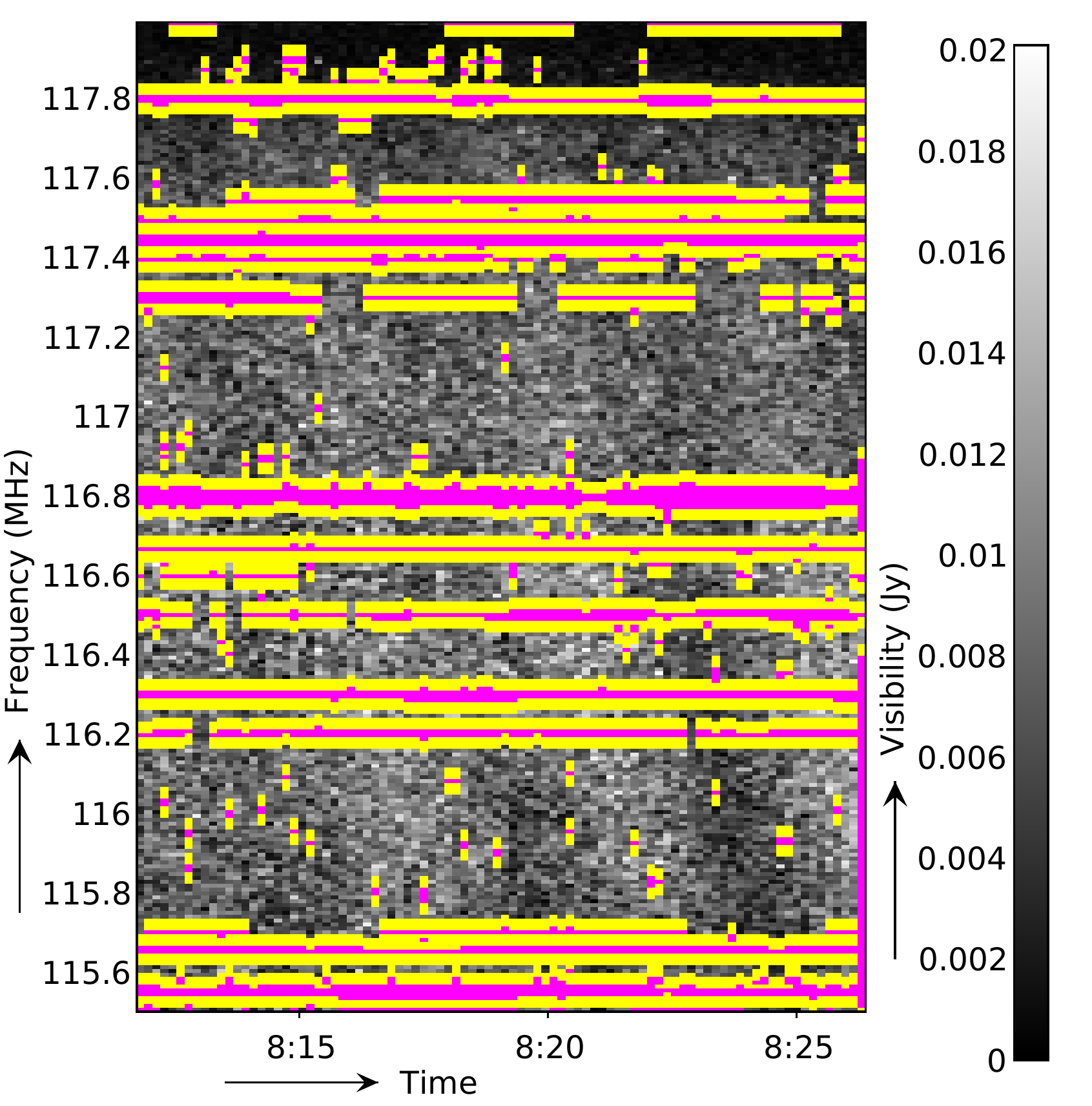}
  }\hspace{3mm}\nolinebreak%
  \subfloat[Dilation with a square kernel]{ \label{fig:wsrt-example-std-dilated-both}
   \includegraphics[width=55mm]{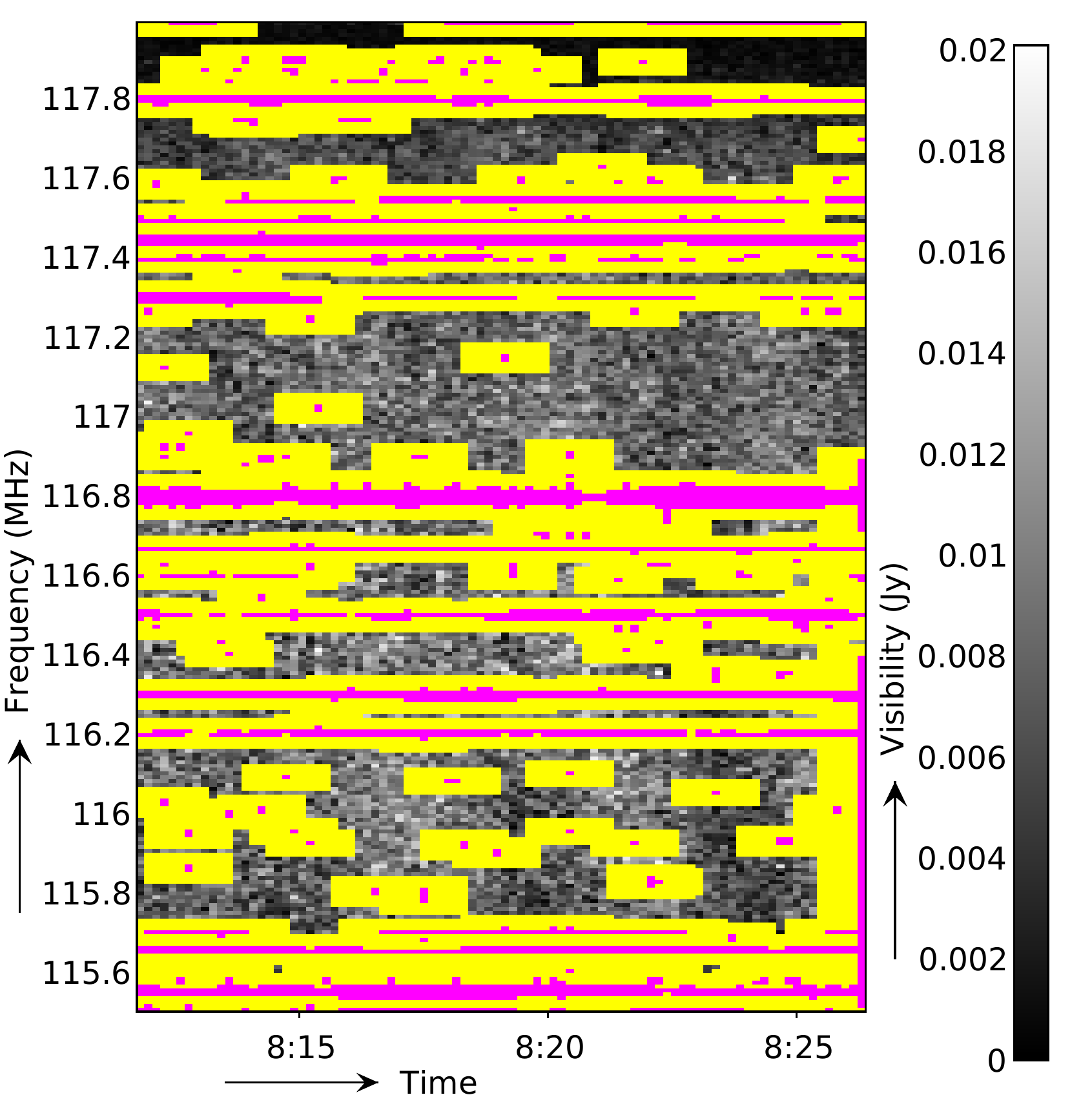}
  }%
 \end{center}
 \caption[]{Gray-scale plots showing examples of the effectiveness of two morphological techniques on the data from Fig.~\ref{fig:wsrt-example}. The pink samples have been set by the \texttt{SumThreshold} algorithm and the yellow samples have additionally been detected with the morphological techniques. Panels~a--c show results of the SIR operator with $\eta=0.2$ in the time direction and/or $\eta=0.3$ in frequency direction. Panels~d--f show an ordinary dilation with a horizontal kernel of five pixels and/or a vertical kernel of three pixels.}
 \label{fig:wsrt-example-dilated}
\end{figure*}

The performance of the SIR operator was tested by using receiver operating characteristics (ROC) analysis. To do so, a ground truth needs to be available, which can only be accurately acquired in a simulated environment. As discussed previously, a very large fraction of RFI is line-like. The samples on such a line are not uniform due to intrinsic effects or instrumental gain variations. Therefore, we have used simulations of four line-shaped RFI features as displayed in Fig.~\ref{fig:features}: (a) a single Gaussian that reaches its $3\sigma$ point at both borders and is 1 in the centre; (b) three periods of a sinusoidal function which is scaled between zero and one; (c) the Gaussian feature, but slanted by 1/50 fraction; and (d) a burst-like signal in which the amplitude levels are drawn from a Rayleigh distribution with mode $\sigma=0.6$. All features are three samples wide. Complex Gaussian distributed noise with $\sigma=1$ was added to the image, such that the amplitudes are Rayleigh distributed. The created two dimensional image of size $180 \times 1024$ was subsequently flagged by the \texttt{SumThreshold} method with settings as in the LOFAR AOFlagger pipeline, and the created flag mask was used as input.

To estimate the performances, the true and false positives ratios (TP and FP ratios respectively) were calculated after detection. We created a fuzzy ground truth mask in which a value of one corresponds with maximal RFI, zero corresponds with samples not contaminated by RFI, and values in between correspond with lower levels of RFI contamination. Fig.~\ref{fig:gaussian-broadband}\subref{fig:gaussian-broadband-no-noise} shows for example the ground-truth mask of the Gaussian feature. Given a sample with ground truth value $\beta$, if the corresponding sample was flagged by the method, it would be counted with ratio $\beta$ as a true positive and $1-\beta$ as a false positive. The total TP and FP ratios are the sum of all the TP and FP values, divided by the total sum of positives and negatives in the test set, respectively.

The SIR operator and a standard morphological dilation have been applied in the direction of the feature, i.e., vertical/frequency direction. The true and false positives were varied by changing the parameter $\eta$ or the dilation size for respectively the SIR operator and the dilation. Different runs gave slightly different results because of the introduced Gaussian noise, hence the simulation was repeated 100 times and the results were averaged.

Figure~\ref{fig:roc-analysis} shows the average results. The shadowed areas in panels~\subref{fig:roc-analysis-rank-operator} and \subref{fig:roc-analysis-dilation} show the standard deviation over the 100 runs. In the case of the Gaussian RFI feature, the \texttt{SumThreshold} pre-detection removes on average 91.3\% RFI power, while simultaneously falsely flagging a ratio of 0.38\%. Hence, if the methods do not flag any additional samples, they have a TP/FP ratio of 91.3\%/0.38\%, and this is therefore the start of both ROC curves for this RFI feature. With $\eta=0.48$, the SIR operator flags all the RFI features with 100\% TP with a FP ratio of 1.36\%, with the exception of the slanted feature. The SumThreshold pre-detection, dilation and SIR-operator work less well on the slanted feature, and fail to detect it with 100\% even at very high sensitivity. The dilation operator needs a size of 32.8\% of the height of the image to detect the vertical RFI features. Since it will dilate any falsely detected input sample equally, its FP ratio is 46\% with this setting. Changing the signal-to-noise ratio (SNR) of the features changes the scaling of the ROC curves, but the relative difference between the two methods remains the same.

Given the various types of RFI, Fig.~\ref{fig:roc-analysis} shows that (I) the SIR operator is extremely accurate on straight features, by detecting all previously undetected samples with only a very slight increase in false positives; (II) the SIR operator is superior to the dilation in all tested situations; and (III) a setting of $\eta\sim0.2$--$0.4$ seems to be a good compromise between FP and TP ratios.

\begin{figure*}
 \begin{center}
\subfloat[Original data]{%
\includegraphics[width=90mm]{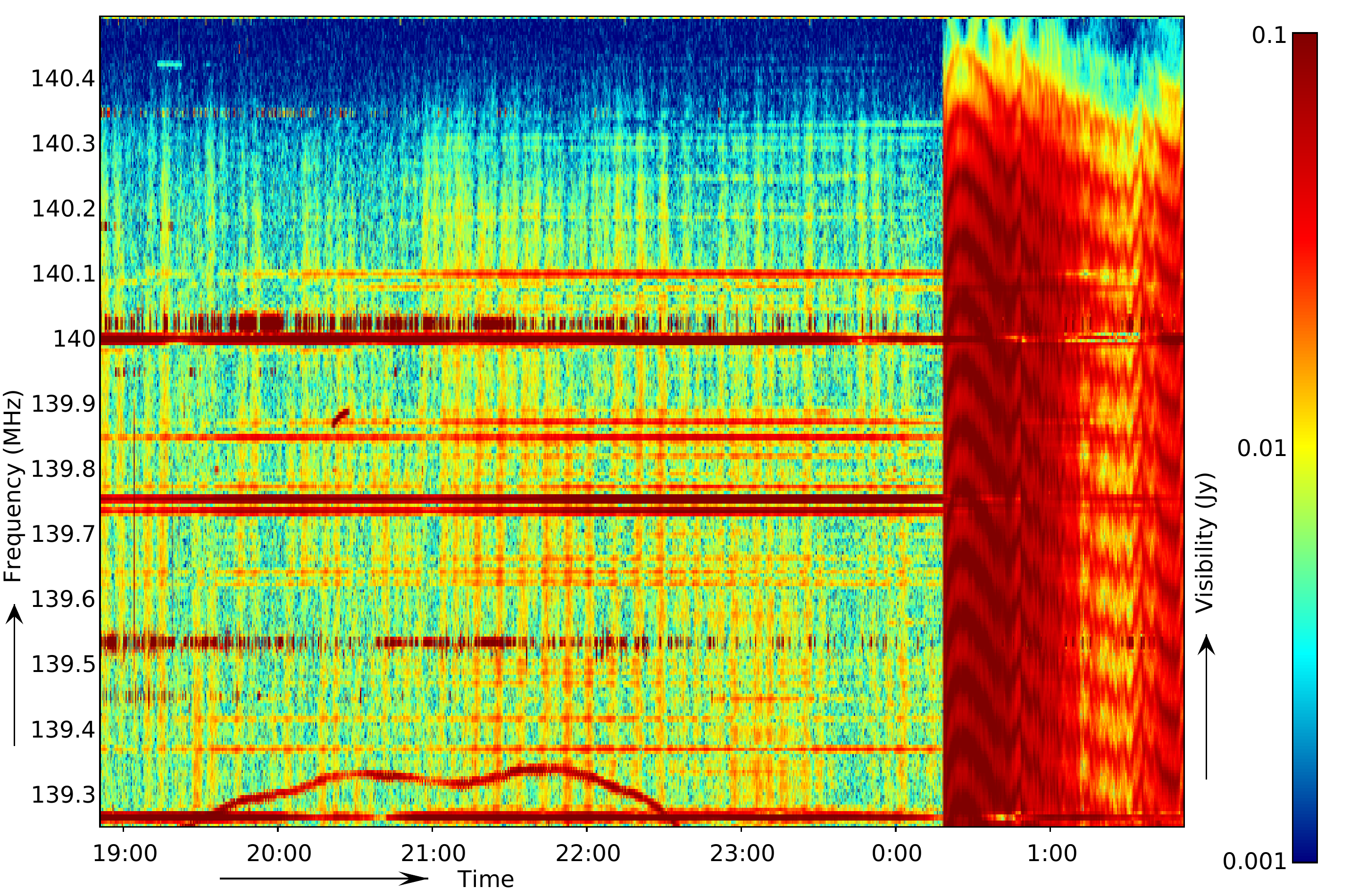}%
}
\subfloat[Flagged with \texttt{SumThreshold} (pink) followed by the SIR\newline operator (yellow)]{%
\includegraphics[width=90mm]{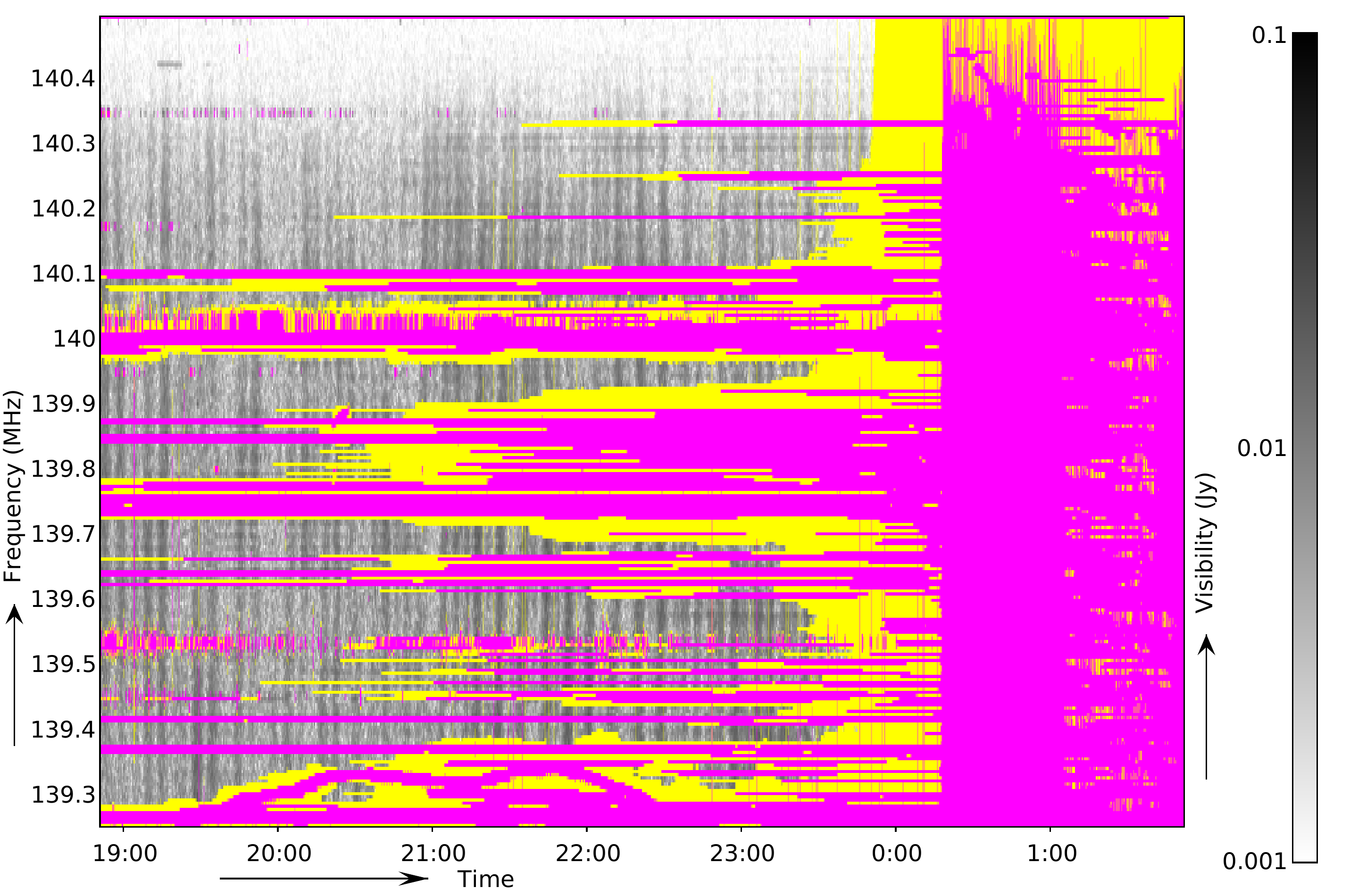}%
}
 \end{center}
 \caption[]{Example of an interesting but uncommon WSRT case: part of a baseline of an observation at 140 MHz that suffered unusually strong broadband RFI during the last 1.5 hrs. It also contains many different kinds of transmitters that mostly occupy constant channels. The vertical stripes are fringes of celestial sources, hence contain the information of interest. The image shown is $2000 \times 250$ samples in size.}
 \label{fig:lots-of-rfi-example}
\end{figure*}

These tests have been performed by applying the operators in one dimension. When applied in two dimensions by using the output of the first dimension as input for the second, the comparison between the dilation and the SIR operator will diverge even more, because the false positives created by the first dimension will be multiplied by the repeated application towards the second dimension. An example of a two-dimensional application is shown in Fig.~\ref{fig:wsrt-example-dilated}. Certain RFI sources create more complex shapes in the time-frequency domain, and contaminate larger non-line like areas. These RFI sources cause higher values in the output of the Gaussian smoothing, which is commonly part of the earlier RFI detection stage, and consequently some of the lower RFI levels of the RFI feature are not flagged. We have seen that the SIR operator will work very well on such features, because it fills the feature and slightly extends the flags in all directions.

It should be noted that one of the assumptions made for the SIR operator to improve detection accuracy, is that parts of the RFI features are not detectable by amplitude thresholding. In practice, however, a small subset of received RFI sources does contribute to an observation with sufficient strength to detect the entire feature with amplitude thresholding. Such transmitters are the worst-case situation for the SIR operator, as the operator will enlarge the flag mask relative to its length, but any samples it flags extra are false positives. Note that it is not useful to perform ROC analysis of such a situation, as the true positives will be constant. The number of false positives can easily be calculated, and scales linearly with $\eta$ and the duration of the transmitter. For example, when applying the SIR operator with $\eta=0.2$ on a strong RFI transmitter that occupies ten minutes of data in one channel, the operator will falsely flag two minutes of the channel before and after it. An example of a band that contains intermittent transmitters is the air traffic communication band of \mbox{118--137~MHz}. Nevertheless, while some of these transmitters are indeed strong, e.g., when they fly through a beam sidelobe, there are also many transmitters at this frequency that are too weakly received to be detected all of the time. Consequently, some of them are only partially detected with amplitude thresholding. This is why we expect better results using the SIR operator even in these bands, compared to using a dilation.

Figure~\ref{fig:lots-of-rfi-example} shows a WSRT example that contains many different RFI kinds. The initial \texttt{SumThreshold} method detects the RFI quite accurately, but it leaves some parts of the last 1.5 hours unflagged. This is solved by the SIR operator, although, because of the sudden start of the RFI, it falsely flags about 20 minutes of data before the start of the RFI. The strong RFI produced by the sporadic transmitter around 140 MHz is flagged by the \texttt{SumThreshold} method, but in this case it is likely that these channels have been occupied all of the time. Therefore, the SIR operator gives the desired result by increasing the flags in those channels. All in all, the baseline might be somewhat overflagged. Nevertheless, it does allow further data reduction without manual intervention, and without thresholding part of the noise. Moreover, this case is exceptional, and the sudden start of very strong broadband RFI is (fortunately) seldomly seen, while the sporadic transmitters such as the one at 140 MHz are seen very often.

A final remark on the ROC analysis performed here is that the given absolute true/false positive ratios are not an accurate representation of actual RFI detection, because our two models are very simplistic and based on the assumption that RFI behaves in a well defined manner. Establishing absolute true/false positive ratios would require a detailed statistical model of the behaviour of RFI. A realistic estimate for the number of samples occupied by RFI with LOFAR is in the order of a few percent \citep{LOFAR-RFI-detection-results}. 

\section{Conclusions and discussion} \label{sec:conclusions}
From panel~\ref{fig:roc-analysis}\subref{fig:roc-analysis-roc} it is clear that the SIR operator is much more suitable to detect the tested kinds of RFI than an ordinary morphological dilation. A value of $\eta=0.2$ was determined by tweaking and validating the results to be a reasonable setting for the LOFAR RFI pipeline, and has been used in the default LOFAR pipeline for over a year. Panel~\ref{fig:roc-analysis}\subref{fig:roc-analysis-rank-operator} shows that this agrees approximately with the simulations: at $\eta=0.2$, the vertical features have almost been completely detected by the SIR operator (Gaussian: 98.9\%, a 7.6\% increase, sinusoidal: 99.9\%, 5.8\% increase, burst: 100\%, 6.7\% increase) with a minor increase in the false positive ratio (Gaussian: 0.69\%, an 0.31\% increase, sinusoidal: 0.95\%, 0.42\% increase, burst: 1.3\%, 0.08\% increase). The slanted feature is not as accurately detected (86\%, 6.1\% increase), but the method does enhance the detection. It is hard to give a similar optimal value for the dilation operation, since the false positives scale linearly with the size of the dilation kernel. Therefore, it depends on what is an acceptable loss in terms of the false positives.

In the case of simulated Gaussian broadband features, only 8.7\% of the RFI power was not detected by the \texttt{SumThreshold} method. For the sinusoidal and burst features, the \texttt{SumThreshold} method performs even better. Therefore, the total benefit of the SIR operator might seem small. However, we think that there are strong reasons to use the method:
\begin{itemize}
\item The added false positives are almost negligible, and the chances of biasing your data are much smaller compared to using amplitude thresholding exclusively. For example, thresholding biases the final distribution of uncorrelated white noise, while morphologically extending a flag mask does not. For these reasons, it is preferable to use morphology to find the final few RFI samples, compared to lowering the threshold.
\item The method is extremely fast and simple, and its processing time is almost negligible in a full RFI pipeline.
\item We have seen situations in which even the low ratio of false negatives that are leaked through an amplitude-based RFI detection pipeline can cause calibration to fail. Empirically, we have seen an improvement of the calibratability of LOFAR observations by using the morphological method.
\end{itemize}

Section \ref{sec:accuracy-analysis} describes that strong intermittent (on $\sim$minute scale) RFI transmitters are probably the worst case for the SIR operator, as in these cases the application of the operator with $\eta=0.2$ could in theory yield 40\% false positives. However, because of LOFAR's high resolution, in combination with the \texttt{SumThreshold}'s unprecedented detection accuracy, the total percentage of flagged data in the case of LOFAR is only a few percent. This implies that even if a large ratio of these were strong intermittent transmitters -- which is unlikely -- the benefits of not having to manually consider data quality in cases where the technique does help, probably outweigh the $\sim$1\% added false positives. If it turns out that some bands do have mostly strong transient transmitters, the $\eta$ parameter could become a function of frequency. At the moment, application of the SIR operator seems to be helpful at any frequency.

In this paper we have assumed scale-invariant behaviour for RFI. In reality this might not be entirely accurate, so instead of using a threshold that grows linearly with the scale, as in our definition of the SIR operator, it might be better to have a threshold that depends on the scale in a non-linear fashion. Also, when looking at the problem from a statistical point of view, RFI might not be equally likely to occur on all scales. For example, RFI might be less likely to occur on a scale of days than on a scale of seconds. When it does occur on large scales, it is doubtful that we actually need to extend the detected intervals in a scale-invariant manner, because the signal would likely already be detected at a smaller scales and gaps would likely be filled. Such considerations would suggest that it might be better to have a threshold that grows less than linearly for large scales. Better RFI statistics and RFI modelling might provide the required information for assessing such considerations.

Several options are available to apply the SIR operator on a two-dimensional input. As shown in Fig.~\ref{fig:lofar-rank-example-2dim}, the intersection of the results in both directions does not extend line RFI, thus is not useful in this context. A union does extend such RFI, but does not extend the flags diagonally. Processing the directions sequentially might therefore be beneficial for RFI that has structure in both frequency and time, as this kind of RFI does likely also slightly contribute in the diagonal direction. The difference between processing time first, frequency first or taking the union of both, is small. Taking the union overcomes the somewhat arbitrary decision of which direction to process first. In the case of LOFAR, we decided to only perform filtering time first, because taking the union of both time and frequency first is more expensive.

Morphology can be used in several image processing tasks, for example in feature detection. Often, generic morphological operations need to be applied on different resolutions. In such cases, the scale-invariant operation to extend binary masks as presented here might be generally useful.

So far, we have considered combinations of one-dimensional application of the SIR operator in order to use it for our two-dimensional application in the time-frequency domain. For this application, but also for more generic applications, it might be interesting to consider a true two-dimensional version of the SIR operator. While the one-dimensional operator selects all subsequences (lines) with a ratio $\ge \eta$ of flagged values, a two-dimensional operator would select all \emph{rectangles} that have a ratio $\ge \eta$ of flagged values. It is however likely that such an operator can not be implemented with a linear time complexity, which makes it less attractive for the large data rate of LOFAR.

We have shown that even slightly slanted features are harder to detect accurately. Fortunately, in the case of LOFAR, such features are very rare. If the features to be detected have a known orientation that is not parallel to one of the axes, it might be an option to apply the operator in the direction of the features. While a trivial implementation can apply the operator along fixed lines, some work might be necessary to maintain translation invariance \citep{directional-morphological-filtering}.
\begin{acknowledgements}
We thank Ger de Bruyn for providing the data and for various discussions. We also thank the LOFAR collaboration for providing the data on which the AOFlagger, including the SIR operator, was developed and tested.
\end{acknowledgements}
\bibliographystyle{aa}
\bibliography{article-rfi-morphology}
\end{document}